\documentclass[ALICE,manyauthors]{cernphprep}
\pdfoutput=1
\usepackage[comma,square,numbers,sort&compress]{natbib}
\usepackage{hyperref}
\usepackage{lineno}
\usepackage{url,color}
\usepackage{enumerate}   
\begin{document}
%%%%%%%%%%%%% ptdr definitions %%%%%%%%%%%%%%%%%%%%%
%
%%%%%%%%%%%%%%%  Title page %%%%%%%%%%%%%%%%%%%%%%%%
%
\newcommand{\PbPb}{\textnormal{Pb-Pb}}
\newcommand{\AuAu}{\textnormal{Au-Au}}
\newcommand{\tn}[1]{\textnormal{#1}}
\newcommand{\snn}{\ensuremath{\sqrt{s_{NN}}}}
\newcommand{\chisqndf}{\chi^2/N_{\rm dof}}
 % off by redefinding CKBNOTEje
\newcommand{\CKBNOTE}[1]{{\bf CKB:  #1}} 
\renewcommand{\CKBNOTE}[1]{}  % switch off

 % off by redefinding RHNOTE
\newcommand{\RHNOTE}[1]{{\bf RH:  #1}} 
\renewcommand{\RHNOTE}[1]{}  % switch off
\newcommand{\multfigsize}{0.95}
\begin{titlepage}
\PHyear{2017}
\PHnumber{215}                 % required, obtained from PH
\PHdate{21 August}              % required, will be obtained from PH
%
%
%%% Put your own title + short title here:
\title{Systematic studies of correlations between different order flow harmonics in Pb-Pb collisions at $\sqrt{s_{NN}}$ = 2.76 TeV}
\ShortTitle{Correlations between different order flow harmonics}   % appears on right page headers

%
%%% Do not change the nexts!
\Collaboration{ALICE Collaboration%
         \thanks{See Appendix~\ref{app:collab} for the list of collaboration
                      members}}
\ShortAuthor{ALICE Collaboration}      % appears on left page headers, do not change
\begin{abstract}
The correlations between event-by-event fluctuations of anisotropic flow harmonic amplitudes
have been measured in Pb-Pb collisions at $\sqrt{s_{NN}}$ = 2.76 TeV with the ALICE detector at the Large Hadron Collider. 
The results are reported in terms of multiparticle correlation observables dubbed symmetric cumulants.
These observables are robust against biases originating from nonflow effects. 
The centrality dependence of correlations between the higher order harmonics (the quadrangular $v_4$ and pentagonal $v_5$ flow) and the lower order harmonics (the elliptic $v_2$ and triangular $v_3$ flow)
is presented. The transverse momentum dependences of correlations between $v_3$ and $v_2$ and between $v_4$ and $v_2$ are also reported. 
The results are compared to calculations from viscous hydrodynamics and  a multiphase transport ({AMPT}) model calculations.
The comparisons to viscous hydrodynamic models demonstrate that
the different order harmonic correlations respond differently to the initial conditions and the temperature dependence of the ratio of shear viscosity to entropy density ($\eta/s$). 
A small average value of $\eta/s$ is favored independent of the specific choice of initial conditions in the models. The calculations with the AMPT initial conditions yield results closest to the measurements. 
Correlations among the magnitudes of $v_2$, $v_3$, and $v_4$ show moderate $p_{\rm T}$ dependence in mid-central collisions. This might be an indication of possible viscous corrections to the equilibrium distribution at hadronic freeze-out, which might help to understand the possible contribution of bulk viscosity in the hadronic phase of the system.
Together with existing measurements of individual flow harmonics, the presented results provide further constraints 
on the initial conditions and the transport properties of the system produced in heavy-ion collisions.

\end{abstract}
\end{titlepage}
\setcounter{page}{2}

% !TEX root = paper.tex

\section{Introduction}
\label{sec:intro}
The main emphasis of the ultrarelativistic heavy-ion collision programs at the Relativistic Heavy Ion Collider (RHIC) and the Large Hadron Collider (LHC) is to study the deconfined phase of strongly interacting QCD matter, the quark-gluon plasma (QGP). 
The matter produced in a heavy-ion collision exhibits strong collective radial expansion~\cite{PhysRevD.34.794,Heinz:2013th}. 
Difference in pressure gradients and the interactions among matter constituents produced in the spatially anisotropic overlap region of the two colliding nuclei result in anisotropic transverse flow in the momentum space.
The large elliptic flow discovered at RHIC energies~\cite{Ackermann:2000tr,Adams:2005dq,Adcox:2004mh,Arsene:2004fa,Back:2004je} is also observed at LHC energies~\cite{Aamodt:2010pa,ALICE:2011ab,Abelev:2014pua,Adam:2016izf,ATLAS:2011ah,ATLAS:2012at,Aad:2013xma,Aad:2014eoa,Chatrchyan:2012wg,Chatrchyan:2012ta,Chatrchyan:2012xq}. The measurements are well described by calculations utilizing viscous hydrodynamics~\cite{Romatschke:2007mq,Shen:2011eg,Schenke:2011zz,Bozek:2012qs,Gale:2012rq,Hirano:2010je}.
These calculations also demonstrated that the shear viscosity to the entropy density ratio ($\eta/s$) of the QGP in heavy-ion collisions at RHIC and LHC energies is close to a universal lower bound $1/4\pi$~\cite{Kovtun:2004de}.

The temperature dependence of $\eta/s$ has some generic features typical to the most known fluids. This ratio reaches its minimum value close to the phase transition region~\cite{Kovtun:2004de,Lacey:2006bc}.
It was shown, using kinetic theory and quantum mechanical considerations~\cite{PhysRevD.31.53}, that $\eta/s\sim0.1$ would be the correct order of magnitude for the lowest possible shear viscosity to entropy density ratio value found in nature. Later it was demonstrated that an exact lower bound $(\eta/s)_{\rm min}=1/4\pi\approx0.08$ can be conjectured using anti-de sitter/conformal field theory (AdS/CFT) correspondence~\cite{Kovtun:2004de}. Hydrodynamical simulations constrained by data support the view that $\eta/s$ of the QGP is close to that limit~\cite{Gale:2012rq}.
It is argued that such a low value might imply that thermodynamic trajectories for the expanding matter would lie close to the quantum chromodynamics (QCD) critical end point, which is another subject of intensive experimental study~\cite{Lacey:2006bc,Csernai:2006zz}.

Anisotropic flow~\cite{Ollitrault:1992bk} is quantified with $n{\mathrm{th}}$-order flow harmonics $v_n$ and corresponding symmetry plane angles $\Psi_n$ in a Fourier decomposition of the particle azimuthal distribution in the plane transverse to the beam direction~\cite{Voloshin:1994mz,Poskanzer:1998yz}:

\begin{equation}
E\frac{{d}^3N}{{d^3}{p}} = \frac{1}{2\pi}\frac{{d}^2N}{p_{\mathrm{T}}{d}p_{\mathrm{T}}{d}\eta} \Big\{1 + 2\sum_{n=1}^{\infty} v_n(p_{\mathrm{T}},\eta) \cos[n(\varphi - \Psi_n)]\Big\},
\label{Eq:Fourier}
\end{equation}

\noindent where $E$, $p$, $p_{\mathrm{T}}$, $\varphi$, and $\eta$ are the particle's energy, momentum, transverse momentum, azimuthal angle, and pseudorapidity, respectively, and $\Psi_n$ is the azimuthal angle of the symmetry plane of the $n{\mathrm{th}}$-order harmonic. Harmonic $v_n$ can be calculated as $v_{n} = \langle{\cos[n(\varphi - \Psi_n)]}\rangle$, where the angular brackets denote an average over all particles in all events.
The anisotropic flow in heavy-ion collisions is typically understood as the hydrodynamic response of the produced matter to spatial deformations of the initial energy density profile~\cite{Floerchinger:2013tya}.
This profile fluctuates event by event due to fluctuating positions of the constituents inside the colliding nuclei, which implies that $v_n$ also fluctuates~\cite{Miller:2003kd,Alver:2006wh}.
The recognition of the importance of flow fluctuations led to the discovery of triangular and higher flow harmonics~\cite{Alver:2010gr,ALICE:2011ab} as well as to the correlations between different $v_{n}$ harmonics~\cite{Niemi:2012aj,Aad:2014fla}.
The higher order harmonics are expected to be sensitive to fluctuations in the initial conditions and to the magnitude of $\eta/s$~\cite{Alver:2010dn,Luzum:2012wu}, while $v_{n}$ correlations have the potential to discriminate between these two respective contributions~\cite{Niemi:2012aj}.

Difficulties in extracting $\eta/s$ in heavy-ion collisions can be attributed mostly to the fact that it strongly depends on the specific choice of the initial conditions in the models used for comparison~\cite{Romatschke:2007mq,Luzum:2012wu,Shen:2011zc}.
Viscous effects reduce the magnitude of the anisotropic flow. Furthermore, the magnitude of $\eta/s$ used in hydrodynamic calculations should be considered as an average over the temperature evolution of the expanding fireball as it is known that $\eta/s$ depends on temperature. 
In addition, part of the anisotropic flow can also originate from the hadronic phase~\cite{Bozek:2011ua,Rose:2014fba,Ryu:2015vwa}. Therefore,
both the temperature dependence of $\eta/s$ and the relative contributions from the partonic and hadronic phases should be understood better to quantify the $\eta/s$ of the QGP.

An important input to the hydrodynamic model simulations is the initial distribution of energy density in the transverse plane (the initial density profile), which is usually estimated from the probability distribution of nucleons in the incoming nuclei.
This initial energy density profile can be quantified by calculating the distribution of the spatial eccentricities $\epsilon_n$~\cite{Alver:2010gr},
\begin{equation}
 \varepsilon_{n} e^{in\Phi_{n}} = -\{r^n e^{in\phi}\}/ \{r^n\},
  \label{eq:eccentricities}
\end{equation}
where the curly brackets denote the average over the transverse plane, i.e. $\{\cdots\}$ = $\int$ $d$$x$$d$$y$\, $e(x,y,\tau_0)$ $(\cdots)$, $r$ is the distance to the system's center of mass, $\phi$ is azimuthal angle, $e(x,y,\tau_0)$ is the energy density at the initial time $\tau_0$, and $\Phi_{n}$ is the participant plane angle (see Refs.~\cite{Teaney:2010vd,Niemi:2015qia}).
There is experimental and theoretical evidence~\cite{Alver:2010gr,Qiu:2011iv,ALICE:2011ab} that the lower order harmonics, $v_2$ and $v_3$, to a good approximation, are linearly proportional to the deformations in the initial energy density in the transverse plane (e.g., $v_n \propto \varepsilon_n$ for $n~=~$2 or 3).
Higher order ($n > 3$) flow harmonics can arise from initial anisotropies in the same harmonic~\cite{Alver:2010gr,Teaney:2010vd,Gubser:2010ui,Hatta:2014jva} (linear response) or can be induced by lower order harmonics~\cite{Bravina:2013xla,Bravina:2013ora} (nonlinear response).
For instance, $v_4$ can develop both as a linear response to $\varepsilon_4$ and/or as a nonlinear response to $\varepsilon_2^2$~\cite{Gardim:2011xv}.
Therefore, the higher harmonics ($n > 3$) can be understood as superpositions of linear and nonlinear responses, through which they are correlated with lower order harmonics~\cite{Teaney:2012ke,Bravina:2013ora,Gubser:2010ui,Hatta:2014jva,Acharya:2017zfg}. When the order of the harmonic is large, the nonlinear response contribution in viscous hydrodynamics is dominant and increases in more peripheral collisions~\cite{Teaney:2012ke,Bravina:2013ora}.
The magnitudes of the viscous corrections as a function of $p_{\rm T}$ for $v_4$ and $v_5$ are sensitive to the ansatz used for the viscous distribution function, a correction for the equilibrium distribution at hadronic freeze-out~\cite{Luzum:2010ad,Teaney:2012ke}.
Hence, studies of the correlations between higher order ($n>3$) and lower order ($v_2$ or $v_3$) harmonics and their $p_{\rm T}$ dependence can help to understand the viscous correction to the momentum distribution at hadronic freeze-out which is among the least understood parts of hydrodynamic calculations~\cite{Dusling:2009df,Teaney:2012ke,Molnar:2014fva,Niemi:2015qia}.

The first results for new multiparticle observables which quantify the relationship between event-by-event fluctuations of two different flow harmonics, the \textit{symmetric cumulants}~(SC), were recently reported by the ALICE Collaboration~\cite{ALICE:2016kpq}.
The new observables are particularly robust against few-particle nonflow correlations~\cite{Aamodt:2010pa} and they provide independent, complementary information to recently analyzed symmetry plane correlators~\cite{Aad:2014fla}. 
It was demonstrated that they are sensitive to the temperature dependence of $\eta/s$ of the expanding medium and therefore simultaneous descriptions of correlations between different order harmonics would constrain both the initial conditions and the medium properties~\cite{ALICE:2016kpq,Zhu:2016puf}.
In this article, we have extended the analysis of SC observables to higher order harmonics (up to fifth order) as well as to the measurement of the $p_{\rm T}$ dependence of correlations for the lower order harmonics ($v_3$-$v_2$ and $v_4$-$v_2$).  We also present a systematic comparison to hydrodynamic and AMPT model calculations.
In Sec.~\ref{sec:method} we present the analysis methods and summarize our findings from the previous work~\cite{ALICE:2016kpq}. The experimental setup and measurements are described in Sec.~\ref{sec:experiment}. The sources of systematic uncertainties are explained in Sec.~\ref{sec:uncertainties}. The results of the measurements are presented in Sec.~\ref{sec:results}. In Sec.~\ref{sec:theory} we present comparisons to model calculations.
Finally, Sec.~\ref{sec:summary} summarizes our new results.
 
\section{Experimental Observables}
\label{sec:method}

Existing measurements for anisotropic flow observables provide an estimate of the average value of $\eta/s$ of the QGP, both at RHIC and LHC energies. What remains uncertain is how the $\eta/s$ of the QGP depends on temperature ($T$). The temperature dependence of $\eta/s$ of the QGP was discussed in Ref.~\cite{Csernai:2006zz}. The effects on hadron spectra and elliptic flow were studied in Ref.~\cite{Niemi:2011ix} for different parametrizations of $\eta/s(T)$.  A more systematic study with event-by-event Eskola-Kajantie-Ruuskanen-Tuominen (EKRT) + viscous hydrodynamic calculations was recently initiated in Ref.~\cite{Niemi:2015qia}, where the first (and only rather qualitative) possibilities were investigated (see Fig.~1 therein). The emerging picture is that the study of individual flow harmonics $v_n$ alone is unlikely to reveal the details of the temperature dependence of $\eta/s$.
It was already demonstrated in Ref.~\cite{Niemi:2015qia} that different $\eta/s(T)$ parametrizations can lead to the same centrality dependence of individual flow harmonics. In Ref.~\cite{Niemi:2012aj} new flow observables were introduced which quantify the degree of correlation between amplitudes of two different harmonics $v_m$ and $v_n$. These new observables have the potential to discriminate between the contributions to anisotropic flow development from initial conditions and from the transport properties of the QGP~\cite{Niemi:2012aj}. Therefore, their measurement would provide experimental constraints on theoretical parameters used to describe the individual stages of the heavy-ion system evolution. In addition, it turned out that correlations of different flow harmonics are sensitive to the temperature dependence of $\eta/s$~\cite{ALICE:2016kpq}, to which individual flow harmonics are weakly sensitive~\cite{Niemi:2015qia}. 
 
For reasons discussed in Refs.~\cite{ALICE:2016kpq,Bilandzic:2013kga}, the correlations between different flow harmonics cannot be studied experimentally with the set of observables introduced in Ref.~\cite{Niemi:2012aj}. 
Based on Ref.~\cite{Bilandzic:2013kga}, new flow observables obtained from multiparticle correlations, \textit{symmetric cumulants}~(SC), were introduced. 

The SC observables are defined as:
 %
%\begin{widetext}
\begin{eqnarray}
\mathrm{SC}(m,n) \equiv \left<\left<\cos(m\varphi_1\!+\!n\varphi_2\!-\!m\varphi_3-\!n\varphi_4)\right>\right>_c &=& \left<\left<\cos(m\varphi_1\!+\!n\varphi_2\!-\!m\varphi_3-\!n\varphi_4)\right>\right>\nonumber\\
&&{}-\left<\left<\cos[m(\varphi_1\!-\!\varphi_2)]\right>\right>\left<\left<\cos[n(\varphi_1\!-\!\varphi_2)]\right>\right>\nonumber\\
&=&\left<v_{m}^2v_{n}^2\right>-\left<v_{m}^2\right>\left<v_{n}^2\right>\,,%\nonumber\\
%&=&0\,.
\label{eq:4p_cumulant}
\end{eqnarray}
%\end{widetext}
%
with the condition $m\neq n$ for two positive integers $m$ and $n$ (for details see Sec.~IV~C in Ref.~\cite{Bilandzic:2013kga}).
In this article, SC($m$,$n$) normalized by the product $\left<v_{m}^2\right>\left<v_{n}^2\right>$~\cite{ALICE:2016kpq,Giacalone:2016afq} is denoted by NSC($m$,$n$):
\begin{equation}
\mathrm{NSC}(m,n) \equiv \frac{\mathrm{SC}(m,n)}{\left<v_{m}^2\right>\left<v_{n}^2\right>}\,.
\label{eq:nsc}
\end{equation}
Normalized symmetric cumulants reflect only the strength of the correlation between $v_{m}$ and $v_{n}$, while SC$(m,n)$ has contributions from both the correlations between the two different flow harmonics and the individual harmonics. In Eq.~(\ref{eq:nsc}) the products in the denominator are obtained from two-particle correlations using a pseudorapidity gap of $|\Delta\eta| > 1.0$ which suppresses biases from few-particle nonflow correlations. For the two two-particle correlations which appear in the definition of SC$(m,n)$ in Eq.~(\ref{eq:4p_cumulant}), the pseudorapidity gap is not needed, since nonflow is suppressed by construction in this observable. This was verified by HIJING model simulations in Ref.~\cite{ALICE:2016kpq}.

The ALICE measurements~\cite{ALICE:2016kpq} have revealed that fluctuations of $v_2$ and $v_3$ are anticorrelated, while fluctuations of $v_2$ and $v_4$ are correlated for all centralities~\cite{ALICE:2016kpq}. It was found that the details of the centrality dependence differ in the fluctuation-dominated (most central) and the geometry-dominated (midcentral) regimes~\cite{ALICE:2016kpq}. The observed centrality dependence of SC(4,2) cannot be captured by models with constant $\eta/s$, indicating that the temperature dependence of $\eta/s$ plays an important role. These results were also used to discriminate between different parametrizations of initial conditions. It was demonstrated that in the fluctuation-dominated regime (central collisions), Monte Carlo (MC)--Glauber initial conditions with binary collision weights are favored over wounded nucleon weights~\cite{ALICE:2016kpq}. 
The first theoretical studies of SC observables can be found in Refs.~\cite{Zhou:2015eya,Giacalone:2016afq,Qian:2016pau,Gardim:2016nrr,Zhu:2016puf,Ke:2016jrd}.

\section{Data Analysis}
\label{sec:experiment}
The data sample of $\PbPb$ collisions at the center-of-mass energy $\snn=2.76$~TeV analyzed in this article was recorded by ALICE during the 2010 heavy-ion run of the LHC.
Detailed descriptions of the ALICE detector can be found
in Refs.~\cite{Aamodt:2008zz,Carminati:2004fp,Alessandro:2006yt}. The time
projection chamber (TPC) was used to reconstruct charged particle
tracks and measure their momenta with full azimuthal coverage in the
pseudorapidity range $|\eta|<0.8$. Two scintillator
arrays (V0A and V0C) which cover the pseudorapidity  ranges $-3.7<\eta<-1.7$
and $2.8<\eta<5.1$ were used for triggering and the determination of
centrality~\cite{Aamodt:2010cz}. The trigger
conditions and the event selection criteria are identical to those
described in Refs.~\cite{Aamodt:2010pa, Aamodt:2010cz}.
Approximately $10^7$ minimum-bias $\PbPb$ events with
a reconstructed primary vertex within $\pm 10$ cm from the nominal
interaction point along the beam direction are selected. Only charged particles reconstructed in the TPC in $|\eta|<0.8$
and $0.2<p_{\rm T}<5$~GeV/$c$ were included in the analysis. The charged track quality cuts
described in Ref.~\cite{Aamodt:2010pa} were applied to minimize
contamination from secondary charged particles and fake tracks.
The track reconstruction efficiency and contamination
were estimated from HIJING Monte Carlo
simulations~\cite{Wang:1991hta} combined with a GEANT3~\cite{Brun:1994aa} detector model and were found to be independent of
the collision centrality. The reconstruction efficiency increases with transverse momenta from
70\% to 80\% for particles with $0.2<p_{\rm T}<1$~GeV/$c$ and remains
constant at $(80 \pm 5)$\% for $p_{\rm T}>1$~GeV/$c$. The estimated
contamination by secondary charged particles from weak decays and
photon conversions is less than 6\% at $p_{\rm T}=0.2$~GeV/$c$ and falls
below 1\% for $p_{\rm T}>1$~GeV/$c$.
The $p_{\rm T}$ cutoff of 0.2~GeV/$c$ reduces event-by-event biases due to small reconstruction efficiency 
at lower $p_{\rm T}$, while the high $p_{\rm T}$ cut-off of 5~GeV/$c$ reduces the effects of jets on the measured correlations. 
Reconstructed TPC tracks constrained to vertex are required to have at least 70 space points (out of a maximum of 159). 
Only tracks with a transverse distance of closest approach to the primary vertex less than 3 mm, both in the longitudinal and transverse directions, are accepted. This reduces the contamination from secondary tracks produced in the detector material, particles from weak decays, etc. Tracks with kinks (i.e., tracks that appear to change direction due to multiple scattering or $K^{\pm}$ decays) were rejected.

\section{Systematic Uncertainties}
\label{sec:uncertainties}
The systematic uncertainties are estimated by varying the event and track selection criteria. All systematic checks described here are performed independently. 
The SC$(m,n)$ values resulting from each variation are compared to ones from the default event and track selection described in the previous section,
and differences are taken as the systematic uncertainty due to each individual source.
The contributions from different sources were added in quadrature to obtain the total systematic uncertainty.

The event centrality was determined by the V0 detectors \cite{Abbas:2013taa} with better than 2\%  resolution for the whole centrality range analyzed. The systematic uncertainty from the centrality determination was evaluated by using the TPC and silicon pixel detector (SPD) \cite{Dellacasa:1999kf} detectors instead of the V0 detectors. 
The systematic uncertainty on the symmetric cumulants which arises from the centrality uncertainty is about 3\% both for SC(5,2) and SC(4,3) and 8\% for  SC(5,3).
As described in Sec.~\ref{sec:experiment}, the reconstructed vertex position along the beam axis ($z$ vertex) is required to be located within 10 cm of the nominal interaction to ensure uniform detector acceptance for tracks within $|\eta|<0.8$. The systematic uncertainty from the $z$-vertex cut was estimated by reducing the $z$-vertex range to 8~cm and was found to be less than 3\%.  

The analyzed events were recorded with two settings of the magnet field polarity and the resulting data sets have almost equal numbers of events. Events with both magnet field polarities were used in the default analysis, and the systematic uncertainties were evaluated from the variation between each of the two magnetic field settings. 
The uncertainty due to the $p_{\rm T}$ dependence of the track reconstruction efficiency was also taken into account.
Magnetic field polarity variation and reconstruction efficiency effects contribute less than 2\% to the systematic uncertainty.

The systematic uncertainty due to the track reconstruction procedure was estimated from comparisons between results for the so-called standalone TPC tracks with the 
same parameters as described in Sec.~\ref{sec:experiment}, and tracks from a combination of the TPC and the inner tracking system (ITS) detectors with tighter selection criteria.
To avoid nonuniform azimuthal acceptance due to dead zones in the SPD, and to get the best transverse momentum resolution, a hybrid track selection utilizing SPD hits and/or ITS refit tracks combined with TPC information was used.
Then each track reconstruction strategy was evaluated by varying the threshold on parameters used to select the tracks at the reconstruction level. 
A systematic difference of up to 12\% was observed in SC$(m,n)$ from the different track selections. 
In addition, we applied the like-sign technique to estimate nonflow contributions~\cite{Aamodt:2010pa} to SC$(m,n)$. The difference between results obtained by selecting all charged particles and results obtained after either selecting only positively or only negatively charged particles was the largest contribution to the systematic uncertainty and is about 7\% for SC(4,3) and 20\% for SC(5,3). 

Another large contribution to the systematic uncertainty originates from azimuthal nonuniformities in the reconstruction efficiency. In order to estimate its effects, we use the AMPT model (see Sec.~\ref{sec:theory}), which has a uniform distribution in azimuthal angle.
Detector inefficiencies were introduced to mimic the nonuniform azimuthal distribution in the data. For the observables SC(5,2), SC(5,3), and SC(4,3), the variation due to nonuniform acceptance is about 9\%, 17\%, and 11\%, respectively.
Overall, the systematic uncertainties are larger for SC(5,3) and SC(5,2) than for the lower harmonics of SC$(m,n)$.
This is because $v_{n}$ decreases with increasing $n$ and becomes more sensitive to azimuthal modulation due to detector imperfections. 

\section{Results}
\label{sec:results}
\begin{figure*}[t!]
            \begin{center}
                       \resizebox{0.51\textwidth}{!}{\includegraphics{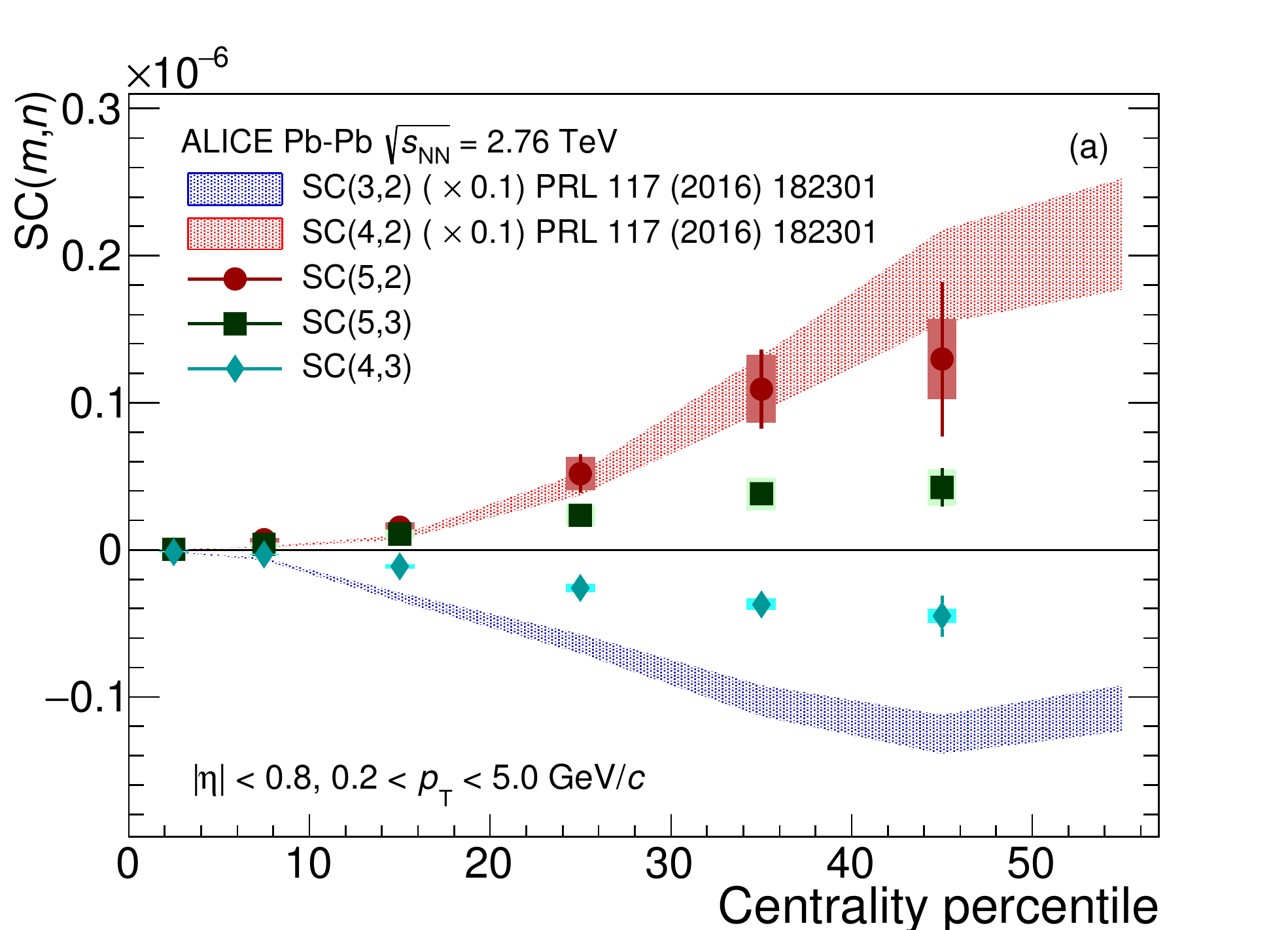}}\hspace{-0.67cm}
                       \resizebox{0.51\textwidth}{!}{\includegraphics{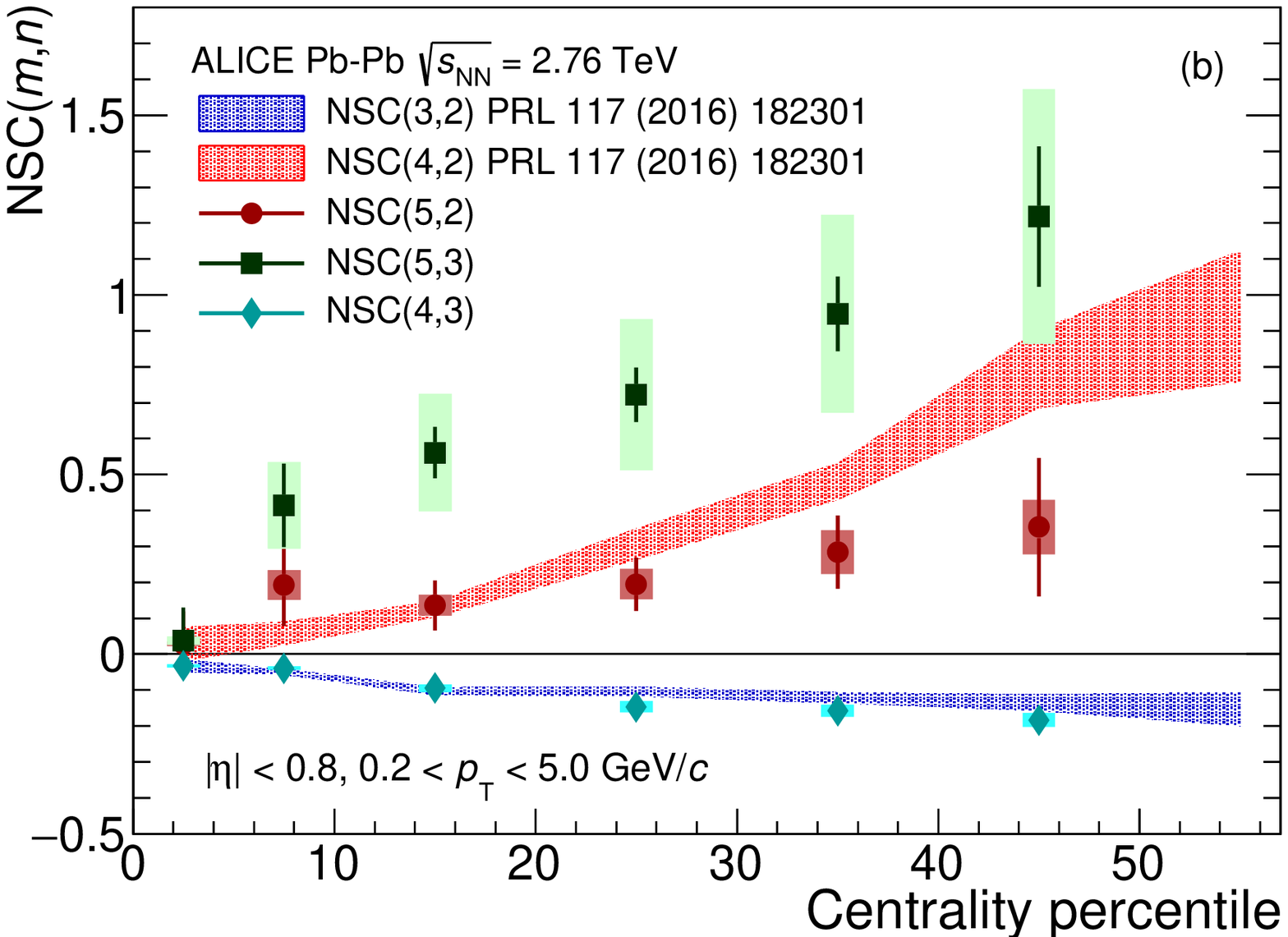}}
        \caption{The centrality dependence of SC($m$,$n$) (a) and NSC($m$,$n$) (b) with flow harmonics for $m$ = 3--5 and $n$ = 2,3 in $\PbPb$ collisions at $\snn=2.76$~TeV. The lower order harmonic correlations [SC(3,2), SC(4,2), NSC(3,2), and NSC(4,2)] are taken from Ref.~\cite{ALICE:2016kpq} and shown as bands. The systematic and statistical errors are combined in quadrature for these lower order harmonic correlations. The SC(4,2) and SC(3,2) are downscaled by a factor of 0.1. Systematic uncertainties are represented with boxes for higher order harmonic correlations.}
        \label{fig:Figure_1}
              \end{center}
\end{figure*}

The centrality dependence of the higher order harmonic correlations [SC(4,3), SC(5,2), and SC(5,3)] are presented in Fig.~\ref{fig:Figure_1} and compared to the lower order harmonic correlations [SC(3,2) and SC(4,2)], which were published in Ref.~\cite{ALICE:2016kpq}. The correlation between $v_3$ and $v_4$ is negative, and similarly for $v_3$ and $v_2$, while the other correlations are all positive, which reveals that $v_2$ and $v_5$ as well as $v_3$ and $v_5$ are correlated like $v_2$ and $v_4$, while $v_3$ and $v_4$ are anticorrelated like $v_3$ and $v_2$.

The higher order flow harmonic correlations are much smaller compared to the lower order harmonic correlations.
In particular, SC(5,2) is 10 times smaller than SC(4,2) and SC(4,3) is about 20 times smaller than SC(3,2).

Unlike SC$(m,n)$, the NSC$(m,n)$ results with the higher order flow harmonics show almost the same order of the correlation strength as the lower order flow harmonic correlations NSC(3,2) or NSC(4,2).
This demonstrates the advantage of using the normalized SC observables in which the correlation strength between flow harmonics is not hindered by the differences in magnitudes of different flow harmonics. The NSC(4,3) magnitude is comparable to NSC(3,2) and one finds that a hierarchy, NSC(5,3) $>$ NSC(4,2) $>$ NSC(5,2), holds for the centrality range 20--50\% within the errors as shown in Fig.~\ref{fig:Figure_1}(b).
The SC(5,2) magnitude is larger than SC(5,3), but the normalized correlation between $v_5$ and $v_3$ is stronger than the normalized correlation between $v_5$ and $v_2$. 
These results indicate that the lower order harmonic correlations are larger than higher order harmonic correlations, not only because of the correlation strength itself but also because of the strength of the individual flow harmonics.

It can be seen in Fig.~\ref{fig:Figure_1}(a) that the lower order harmonic correlations as well as SC(5,2) increase nonlinearly toward peripheral collisions.
In the case of SC(5,3) and SC(4,3), the centrality dependence is weaker than for the other harmonic correlations.
The NSC(5,3) observable shows the strongest normalized correlation among all harmonics while NSC(5,2) shows the weakest centrality dependence.
Both NSC(3,2) and NSC(4,3) are getting more anticorrelated toward peripheral collisions and have similar magnitudes.

\begin{figure}[t!]
	\begin{center}
        	\resizebox{0.92\columnwidth}{!}{\includegraphics{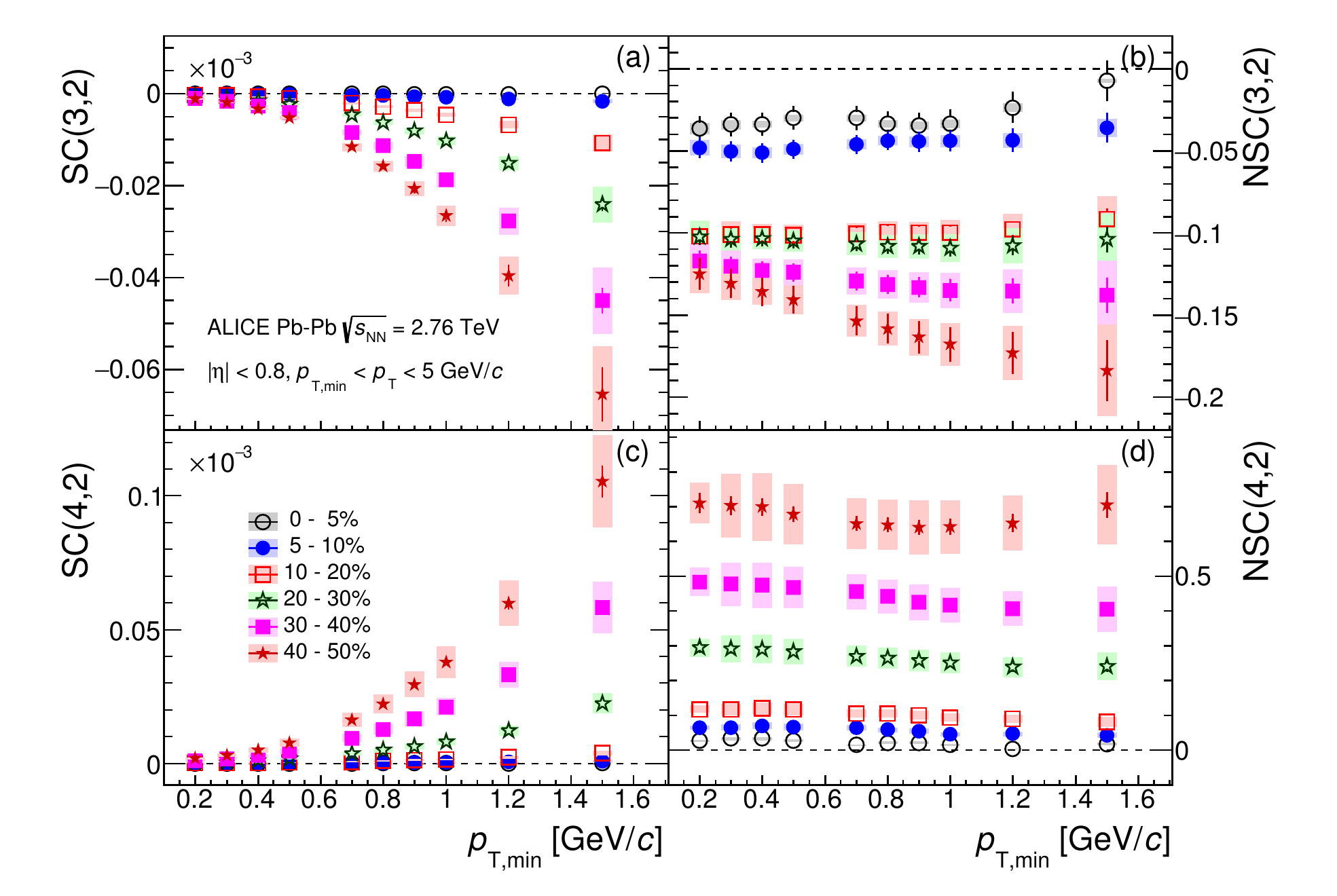}}
        \caption{SC(3,2) and SC(4,2) [panels (a) and (c)] as a function of minimum $p_{\rm T}$ cuts in $\PbPb$ collisions at $\snn=2.76$~TeV are shown in the left panels. The NSC(3,2) and NSC(4,2) [panels (b) and (d)] are shown in the right panels. Systematic uncertainties are represented with boxes.}         \label{fig:Figure_2}
        \end{center}   
\end{figure}

To study the $p_{\rm T}$ dependence of SC$(m,n)$, we present the results as a function of the low $p_{\rm T}$ cutoff ($p_{\rm T, min}$), instead of using independent $p_{\rm T}$ intervals; this decreases large statistical fluctuations in the results. Various minimum $p_{\rm T}$ cuts from 0.2 to 1.5~GeV/$c$ are applied.
The $p_{\rm T}$ dependent results for SC(3,2) and SC(4,2) as a function of minimum $p_{\rm T}$ cuts are shown in Figs.~\ref{fig:Figure_2}(a) and \ref{fig:Figure_2}(c).
The strength of SC$(m,n)$ becomes larger as $p_{\rm T, min}$ increases. 
The centrality dependence is stronger with higher $p_{\rm T, min}$ cuts, with SC$(m,n)$ getting much larger as centrality percentile or $p_{\rm T, min}$ increases. 
The NSC(3,2) and NSC(4,2) observables with different $p_{\rm T, min}$ are shown in Figs.~\ref{fig:Figure_2}(b) and \ref{fig:Figure_2}(d).
The strong $p_{\rm T, min}$ dependence observed in SC$(m,n)$ is not seen in NSC$(m,n)$. 
This indicates that the $p_{\rm T}$ dependence of SC$(m,n)$ is dominated by the $p_{\rm T}$  dependence of the individual flow harmonics $\left<v_n\right>$. 
The $p_{\rm T, min}$ dependence of NSC(3,2) is not clearly seen and it is consistent with no $p_{\rm T, min}$ dependence within the statistical and systematic errors for the centrality range 0--30\%, while showing a moderate increase of anticorrelation with increasing $p_{\rm T, min}$ for the 30--50\% centrality range.
The NSC(4,2) observable shows a moderate decreasing trend as $p_{\rm T, min}$ increases. These observations are strikingly different from the $p_{\rm T}$ dependence of the individual flow harmonics, where the relative flow fluctuations $\sigma_{v_2}/\langle v_{2} \rangle$~\cite{Voloshin:2008dg} are independent of transverse momentum up to $p_{\rm T}$ $\sim$ 8~GeV/$c$ (see Fig. 3 in Ref.~\cite{Abelev:2012di}).

\begin{figure}[t!]
	\begin{center}
        	\resizebox{0.92\columnwidth}{!}{\includegraphics{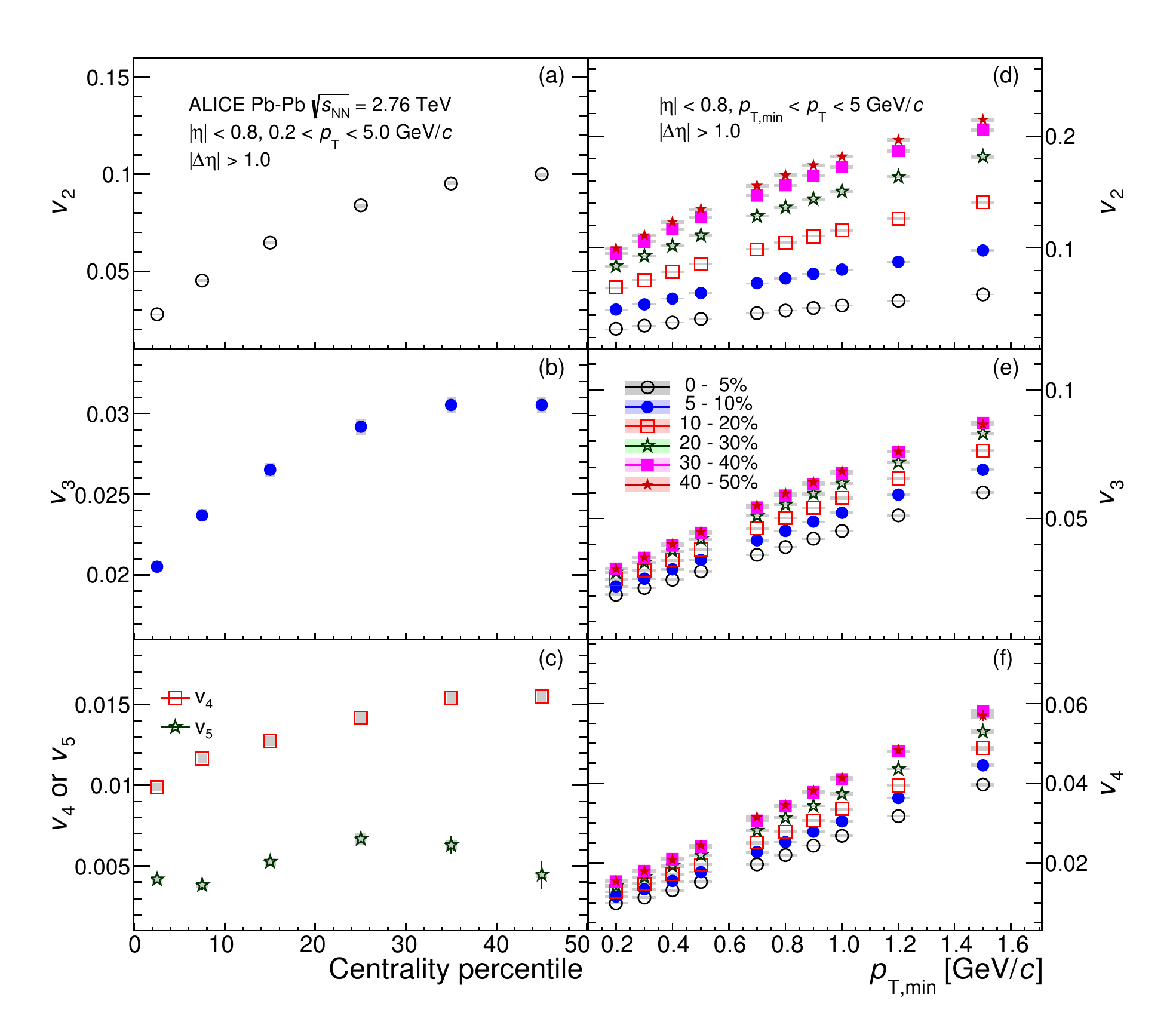}}
        \caption{The individual flow harmonics $v_n$ for $n$ = 2--5 in $\PbPb$ collisions at $\snn=2.76$~TeV are shown in the left panels [(a), (b), and (c)]. $v_4$ and $v_5$ are shown in the same panel (c). The $p_{\rm T, min}$ dependence of $v_n$ for $n$ = 2--4 is shown in the right panels [(d), (e), and (f)].}
        \label{fig:Figure_3}
        \end{center}   
\end{figure}

As discussed in Sec.~\ref{sec:method}, the NSC($m$,$n$) observables are normalized by the product $\left<v_{m}^2\right>\left<v_{n}^2\right>$.
These products are obtained from two-particle correlations using a pseudorapidity gap of $|\Delta\eta| > 1.0$. 
In this paper, we denote the $p_{\rm T}$ integrated $v_{n}\{2, |\Delta\eta|>1\}$ as $v_n$ in the transverse momentum range $0.2<p_{\rm T}<5.0$~GeV/$c$.
The individual flow harmonics $v_n$ used in calculations of the NSC observables are shown in Fig.~\ref{fig:Figure_3}.
The centrality dependence of $v_n$ for $n$ = 2--5 is shown in Figs.~\ref{fig:Figure_3}(a)--\ref{fig:Figure_3}(c). The $v_n$ values ($n < 5$) are equivalent to those in Ref.~\cite{Adam:2016izf}. The fifth-order flow harmonic $v_5$ is shown in Fig.~\ref{fig:Figure_3}(c). The $p_{\rm T, min}$ dependence of $v_n$ for $n$ = 2--4 is shown in Figs.~\ref{fig:Figure_3}(d)--\ref{fig:Figure_3}(f) in all centrality ranges relevant to the measured NSC($m$,$n$) observables.

\section{Model Comparisons}
\label{sec:theory}
We have performed a systematic comparison of the centrality and transverse momentum dependence of the SC($m$,$n$) and NSC($m$,$n$) to the event-by-event  EKRT+viscous hydrodynamics~\cite{Niemi:2015qia}, VISH2+1~\cite{Shen:2010uy,Shen:2014vra}, and the AMPT~\cite{Qian:2016pau,Lin:2004en,Lin:2014tya} models. Comparisons for $v_n$ coefficients with the model calculations are presented in the appendix.

In the event-by-event EKRT+viscous hydrodynamic calculations~\cite{Niemi:2015qia}, the initial energy density profiles are calculated using a next-to-leading order perturbative-QCD + saturation model~\cite{Paatelainen:2012at,Paatelainen:2013eea}. The subsequent space-time evolution is described by relativistic dissipative fluid dynamics with different parametrizations for the temperature dependence of the shear viscosity to entropy density ratio $\eta/s(T)$. 
This model gives a good description of the charged hadron multiplicity and the low-$p_{\rm T}$ region of the charged hadron spectra at RHIC and the LHC (see Figs.~11--13 in Ref.~\cite{Niemi:2015qia}).
Each of the $\eta/s(T)$ parametrizations is adjusted to reproduce the measured $v_n$ from central to midperipheral collisions (see Fig.~15 in Ref.~\cite{Niemi:2015qia} and our appendix). 

The VISH2+1~\cite{Shen:2010uy,Shen:2014vra} event-by-event calculations for relativistic heavy-ion collisions are based on (2+1)-dimensional viscous hydrodynamics which describes the QGP phase and the highly dissipative and off-equilibrium late hadronic stages with fluid dynamics. By tuning transport coefficients and decoupling temperature for a given scenario of initial conditions, it can describe the $p_{\rm T}$ spectra and different flow harmonics at RHIC and the LHC~\cite{Qiu:2011hf, Shen:2010uy, Shen:2011eg, Bhalerao:2015iya} energies.
Three different types of initial conditions~\cite{Zhu:2016puf} ({MC-Glauber}, Monte Carlo Kharzeev-Levin-Nardi ({MC-KLN}), and {AMPT}) along with different constant $\eta/s$ values have been used for our data to model comparisons.
Traditionally, the Glauber model constructs the initial entropy density from the wounded nucleon and binary collision density profiles~\cite{Kolb:2000sd}. The {KLN} model assumes that the initial energy density is proportional to that of the initial gluons calculated from the corresponding $k_{\rm T}$ factorization formula~\cite{Kharzeev:2000ph}.
In Monte Carlo versions {MC-Glauber} and {MC-KLN}~\cite{Miller:2007ri,Drescher:2006ca,Hirano:2009ah} of these models, additional initial state fluctuations are introduced through position fluctuations of individual nucleons inside the colliding nuclei. For the {AMPT} initial conditions~\cite{Bhalerao:2015iya,Pang:2012he,Xu:2016hmp}, the fluctuating energy density profiles are constructed from the energy distribution of individual partons, which fluctuate in both momentum and coordinate space. Compared with the {MC-Glauber} and {MC-KLN} initial conditions, the additional Gaussian smearing in the {AMPT} initial conditions gives rise to nonvanishing initial local flow velocities~\cite{Pang:2012he}. 

Even though thermalization could be achieved quickly in collisions of very large nuclei and/or at extremely high energy~\cite{Kurkela:2015qoa}, the dense matter created in heavy-ion collisions may not reach full thermal or chemical equilibrium due to its finite size and short lifetime. To address such nonequilibrium many-body dynamics, the AMPT model~\cite{Zhang:1999bd,Lin:2000cx,Lin:2004en} has been developed, which includes both initial partonic and final hadronic interactions and the transition between these two phases of matter.
The initial conditions in the AMPT are given by the spatial and momentum distributions of minijets and soft strings from the HIJING model~\cite{Wang:1991hta,Gyulassy:1994ew}.
For the data comparisons, three different configurations of the AMPT model have been used: the default one and string melting with and without hadronic rescattering. The input parameters used in all configurations are $\alpha_s = 0.33$ and a partonic cross section of 1.5~mb.
In the default configuration, partons are recombined with their parent strings when they stop interacting. The resulting strings are later converted into hadrons using the Lund string fragmentation model~\cite{Andersson:1986gw,NilssonAlmqvist:1986rx}.
The Lund string fragmentation parameters were set to $\alpha = 0.5$ and $b = 0.9$~GeV$^{-2}$.
In the string melting configuration, the initial strings are melted into partons whose interactions are described by the Zhang's parton cascade (ZPC) model~\cite{Zhang:1997ej}. These partons are then combined into the final-state hadrons via a quark coalescence model. 
In both configurations, the dynamics of the subsequent hadronic matter is described by a hadronic cascade based on a relativistic transport (ART) model~\cite{Li:2001xh} which includes resonance decays.
The string melting configuration of the AMPT without hadronic rescattering was used to study the influence of the hadronic phase on the development of the anisotropic flow.
Even though the string melting version of AMPT~\cite{Lin:2001zk,Lin:2004en} reasonably well reproduces particle yields, $p_{\rm T}$ spectra, and $v_2$ of low-$p_{\rm T}$ pions and kaons in central and midcentral $\AuAu$ collisions at $\snn=200$~GeV and $\PbPb$ collisions at $\snn=2.76$~TeV~\cite{Lin:2014tya}, it was observed in a recent study~\cite{Adam:2016nfo} that it fails to quantitatively reproduce the flow harmonics of identified hadrons ($v_2$, $v_3$, $v_4$, and $v_5$) at $\snn=2.76$~TeV. It turns out that the radial flow in AMPT is 25\% lower than that measured at the LHC, which is responsible for this quantitative disagreement~\cite{Adam:2016nfo}. The details of the AMPT configurations used in this article and the comparisons of $p_{\rm T}$-differential $v_{n}$ for pions, kaons, and protons to the data can be found in Ref.~\cite{Adam:2016nfo}.

\subsection{Centrality dependence of SC($m$,$n$) and NSC($m$,$n$)}
\label{sec:theory_allorder}
Comparison to event-by-event EKRT+viscous hydrodynamic predictions with various parametrizations of the temperature dependence of $\eta/s(T)$ was shown in Fig.~2 of Ref.~\cite{ALICE:2016kpq}.
It was demonstrated that NSC(3,2) is sensitive mainly to the initial conditions, while NSC(4,2) is sensitive to both the initial conditions and the system properties, which is consistent with the predictions from Ref.~\cite{Niemi:2012aj}.
The model calculations for NSC(4,2) observable show that it has better sensitivity for different $\eta/s(T)$ parametrizations but they cannot
describe either the centrality dependence or the absolute values. The discrepancy between data and theoretical predictions indicates that the current understanding of initial conditions in models of heavy-ion collisions needs to be revisited to further constrain $\eta/s(T)$.
The measurement of SC($m$,$n$) and NSC($m$,$n$) can provide new constraints for the detailed modeling of fluctuating initial conditions.

\begin{figure}[t!]
            \begin{center}
                       \resizebox{\multfigsize\textwidth}{!}{\includegraphics{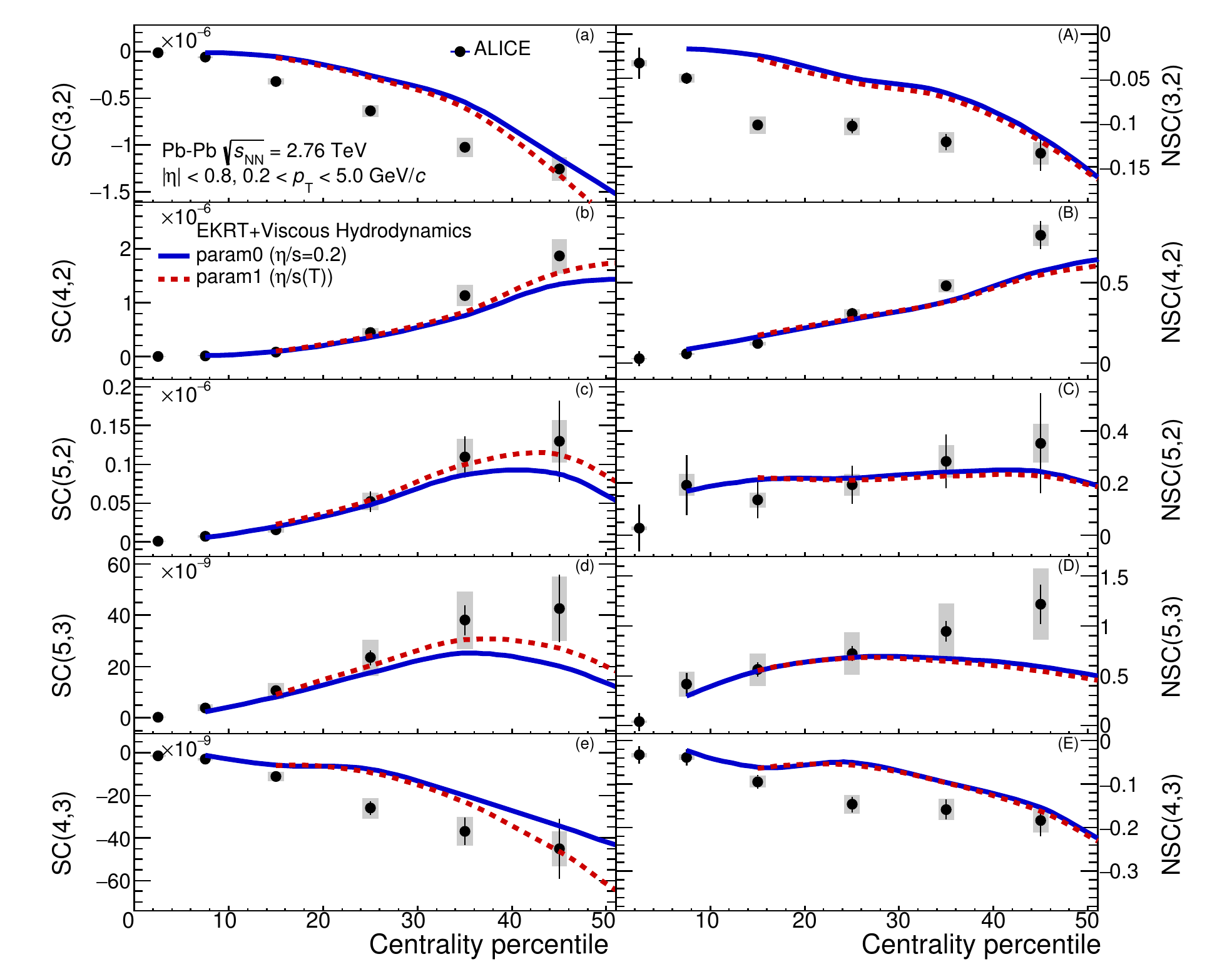}}
              \end{center}
        \caption{The centrality dependence of SC($m$,$n$) and NSC($m$,$n$) in $\PbPb$ collisions at $\snn=2.76$~TeV. Results are compared to the event-by-event EKRT+viscous hydrodynamic calculations~\cite{Niemi:2015qia}. The lines are hydrodynamic predictions with two different $\eta/s(T)$ parametrizations. Left (right) panels show SC($m$,$n$) (NSC($m$,$n$)).}        
        \label{fig:Figure_4}
\end{figure}

The calculations for the two sets of parameters which describe the lower order harmonic correlations best are compared to the data in Fig.~\ref{fig:Figure_4}. 
As can be seen in Fig.~1 from Ref.~\cite{Niemi:2015qia}, for the ``param1'' parametrization the phase transition from the hadronic to the QGP phase occurs at the lowest temperature, around 150~MeV. This parametrization is also characterized by a moderate slope in $\eta/s(T)$ which decreases (increases) in the hadronic (QGP) phase.
The model calculations in which the temperature of the phase transition is larger than for ``param1'' are ruled out by the previous measurements~\cite{ALICE:2016kpq}.
While the correlations between $v_5$ and $v_2$ are well described at all centralities, the correlations between $v_5$ and $v_3$ are reproduced in the 0--40\% centrality range and deviate by about one $\sigma$ for 40--50\% centrality.
In the case of $v_4$ and $v_3$, the same models underestimate the anticorrelation in the data significantly in midcentral collisions and fail similarly for the anticorrelation between $v_3$ and $v_2$.

\begin{figure}[t!]
	\begin{center}
        	\resizebox{\multfigsize\columnwidth}{!}{\includegraphics{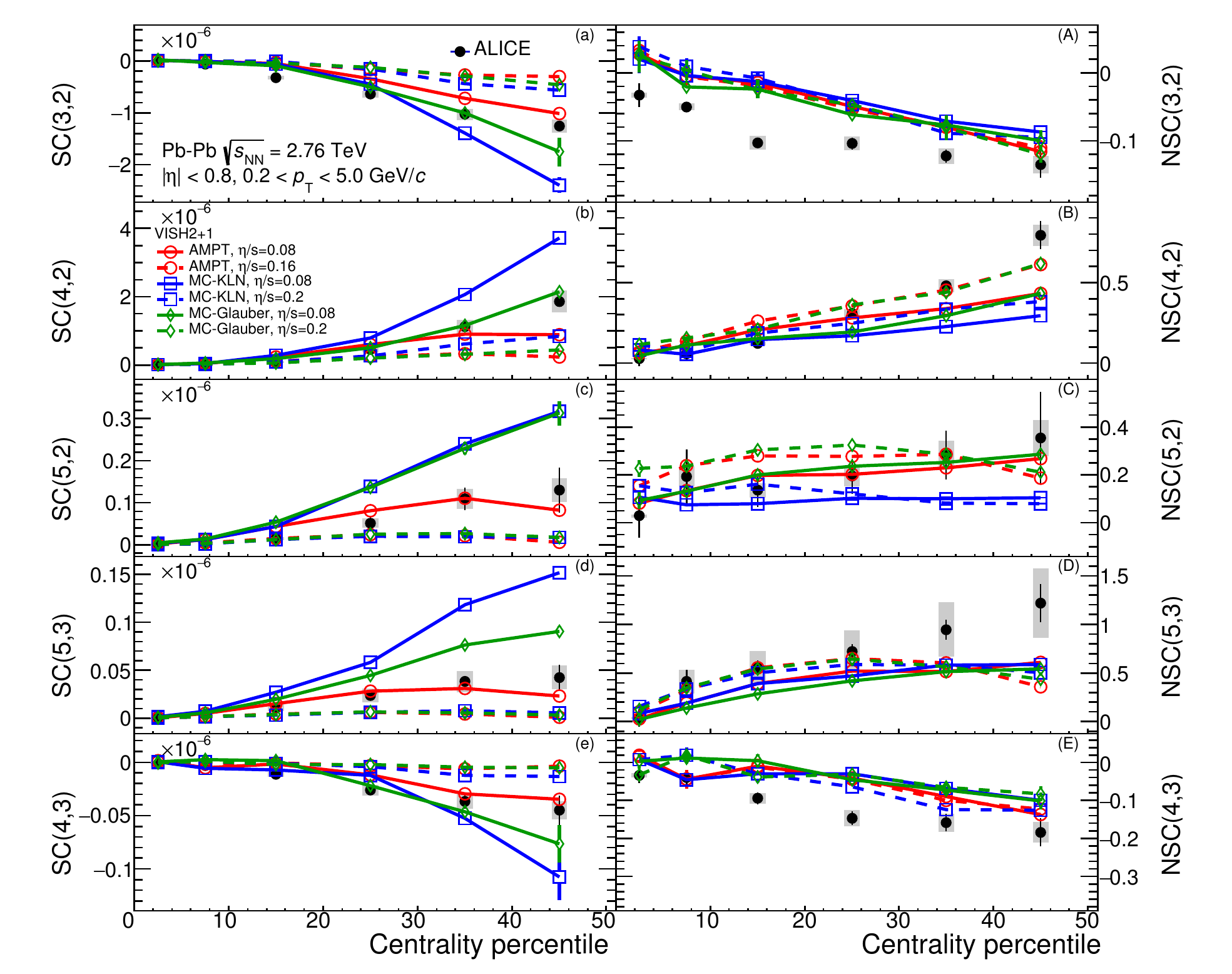}}
        \caption{The centrality dependence of SC($m$,$n$) and NSC($m$,$n$) in $\PbPb$ collisions at $\snn=2.76$~TeV. Results are compared to various VISH2+1 calculations~\cite{Zhu:2016puf}. Three initial conditions from AMPT, MC-KLN, and MC-Glauber are drawn as different colors and markers. The $\eta/s$ parameters are shown as different line styles, the small shear viscosity ($\eta/s=0.08$) are shown as solid lines, and large shear viscosities ($\eta/s=0.2$ for MC-KLN and MC-Glauber and 0.16 for AMPT) are drawn as dashed lines. Left (right) panels show SC($m$,$n$)  (NSC($m$,$n$)).}
        \label{fig:Figure_5}
        \end{center}   
 \end{figure}
 
The comparison to the VISH2+1 calculation~\cite{Zhu:2016puf} is shown in Fig.~\ref{fig:Figure_5}.  All calculations with large $\eta/s$ regardless of the initial conditions ($\eta/s=0.2$ for MC-KLN and MC-Glauber initial conditions and $\eta/s=0.16$ for AMPT initial conditions) fail to describe the centrality dependence of the SC($m$,$n$) observables of all orders, shown in the left panels in Fig.~\ref{fig:Figure_5}.
Among the calculations with small $\eta/s$ ($\eta/s=0.08$), the one with the AMPT initial conditions describes the data better than the ones with other initial conditions for all SC($m$,$n$) observables measured, but it cannot describe the data quantitively for most of the centrality ranges.

However, NSC(4,2) is sensitive both to the initial conditions and the $\eta/s$ parametrizations used in the models.
Even though NSC(4,2) favors both AMPT initial conditions with $\eta/s=0.08$ and MC-Glauber initial conditions with $\eta/s=0.20$,
SC(4,2) can only be described by models with smaller $\eta/s$. Hence the calculation with large $\eta/s=0.20$ is ruled out. We conclude that $\eta/s$ should be small and that AMPT initial conditions are favored by the data.
The NSC(5,2) and NSC(5,3) observables are quite sensitive to both the initial conditions and the $\eta/s$ parametrizations.
The SC(4,3) results clearly favor smaller $\eta/s$ values but NSC(4,3) cannot be described by these models quantitively.

 \begin{figure}[t!]
	\begin{center}
        	\resizebox{\multfigsize\columnwidth}{!}{\includegraphics{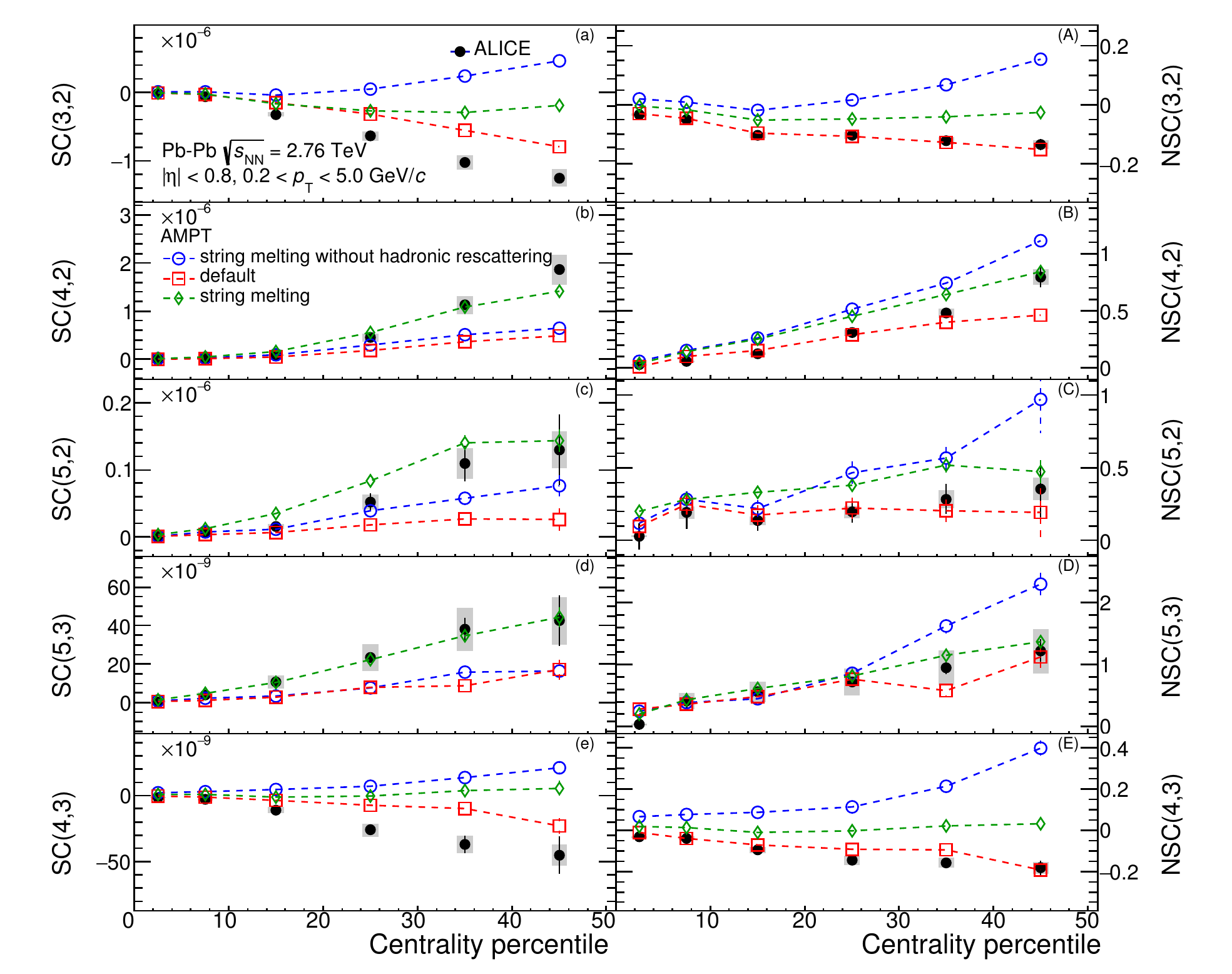}}
        \caption{The centrality dependence of SC($m$,$n$) and NSC($m$,$n$) in $\PbPb$ collisions at $\snn=2.76$~TeV. Results are compared to various AMPT models. Left (right) panels show SC($m$,$n$)  (NSC($m$,$n$)).}
        \label{fig:Figure_6}
        \end{center}   
 \end{figure}
 
The SC($m$,$n$) and NSC($m$,$n$) observables calculated from AMPT simulations are compared with data in Fig.~\ref{fig:Figure_6}.
For SC(3,2), the calculation with the default AMPT settings is closest to the data, but none of the AMPT configurations can describe the data fully. 
The third version based on the string melting configuration without the hadronic rescattering phase is also shown.
The hadronic rescattering stage makes both SC(3,2) and NSC(3,2) smaller in the string melting AMPT model but not enough to describe the data.
Further investigations proved why the default AMPT model can describe NSC(3,2) but underestimates SC(3,2). By taking the differences in the individual flow harmonics ($v_2$ and $v_3$) between the model and data into account, it was possible to recover the difference in SC(3,2) between the data and the model. The discrepancy in SC(3,2) can be explained by the overestimated individual $v_n$ values as reported in Ref.~\cite{Adam:2016nfo} in all centrality ranges. 

In the case of SC(4,2), the string melting configuration of the AMPT model can describe the data fairly well while the default configuration underestimates it.
The NSC(4,2) observable is slightly overestimated by the string melting setting which can describe SC(4,2) but the default AMPT configuration can describe the data better.
The influence of the hadronic rescattering phase on NSC(4,2) is opposite to other observables [SC(3,2), NSC(3,2), and SC(4,2)]. The hadronic rescattering makes NSC(4,2) slightly smaller.
It should be noted that the agreement with SC($m$,$n$) should not be overemphasized since there are discrepancies in the individual $v_n$ between the AMPT models and the data as was demonstrated for SC(3,2).
Hence, the simultaneous description of SC($m$,$n$) and NSC($m$,$n$) should give better constraints on the parameters in AMPT models.
The string melting AMPT model describes SC(5,3) and NSC(5,3) well. However, the same setting overestimates SC(5,2) and NSC(5,2). 
The default AMPT model can describe NSC(5,3) and NSC(5,2) fairly well, as in the case of NSC(3,2) and NSC(4,2).
In the case of SC(4,3), neither of the settings can describe the data but the default AMPT model comes the closest to the data. 
The NSC(4,3) observable is well described by the default AMPT model but cannot be reproduced by the string melting AMPT model.
In summary, the default AMPT model describes well the normalized symmetric cumulants [NSC($m$,$n$)] from lower to higher order harmonic correlations while the string melting AMPT model overestimates NSC(3,2) and NSC(5,2) and predicts a very weak correlation both for NSC(3,2) and NSC(4,3). 

As discussed in Sec.~\ref{sec:results}, a hierarchy NSC(5,3) $>$ NSC(4,2) $>$ NSC(5,2) holds for centrality ranges $>~20\%$ within the errors.
Except for the 0--10\% centrality range, we found that the same hierarchy also holds in the hydrodynamic calculations and the AMPT models explored in this article.
While NSC(5,2) is smaller than NSC(5,3), SC(5,2) is larger than SC(5,3).
The observed inverse hierarchy, SC(5,2) $>$ SC(5,3), can be explained by different magnitudes of the individual flow harmonics ($v_2$ $>$ $v_3$). 
This can be attributed to the fact that flow fluctuations are stronger for $v_3$ than $v_2$~\cite{Aad:2013xma}. This was claimed in Ref.~\cite{Zhu:2016puf} and also seen in Ref.~\cite{Bhalerao:2014xra} based on the AMPT model calculations. 
NSC($m$,$n$) correlators increase with larger $\eta/s$ in hydrodynamic calculations in the 0--30\% centrality range in the same way as the event plane correlations~\cite{Bhalerao:2013ina,Teaney:2013dta}. In semiperipheral collisions ($>$~40\%), the opposite trend is observed.

We list here the important findings from the model comparisons to the centrality dependence of SC($m$,$n$) and NSC($m$,$n$):
\begin{enumerate}[(i)]
	\item The NSC(3,2) observable is sensitive mainly to the initial conditions, while the other observables are sensitive to both the initial conditions and the temperature dependence of $\eta/s$.
	\item The correlation strength between $v_3$ and $v_2$ and between $v_4$ and $v_3$ [SC(3,2), SC(4,2), NSC(3,2), and NSC(4,3)] is significantly underestimated in hydrodynamic model calculations in midcentral collisions.
	\item All the VISH2+1 model calculations with large $\eta/s$ fail to describe the centrality dependence of the correlations regardless of the initial conditions.
	\item Among the VISH2+1 model calculations with small $\eta/s$ ($\eta/s=0.08$), the one with the AMPT initial conditions describes the data qualitatively but not quantitively for most of the centrality ranges.
	\item The default AMPT model can describe the normalized symmetric cumulants [NSC($m$,$n$)] quantitively for most centralities while the string melting AMPT model fails to describe them.
	\item A hierarchy NSC(5,3) $>$ NSC(4,2) $>$ NSC(5,2) holds for centrality percentile ranges $>~20\%$ within the errors. This hierarchy is reproduced well both by hydrodynamic and AMPT model calculations.
\end{enumerate}

\begin{figure}[t!]
            \begin{center}
                       \resizebox{0.99\textwidth}{!}{\includegraphics{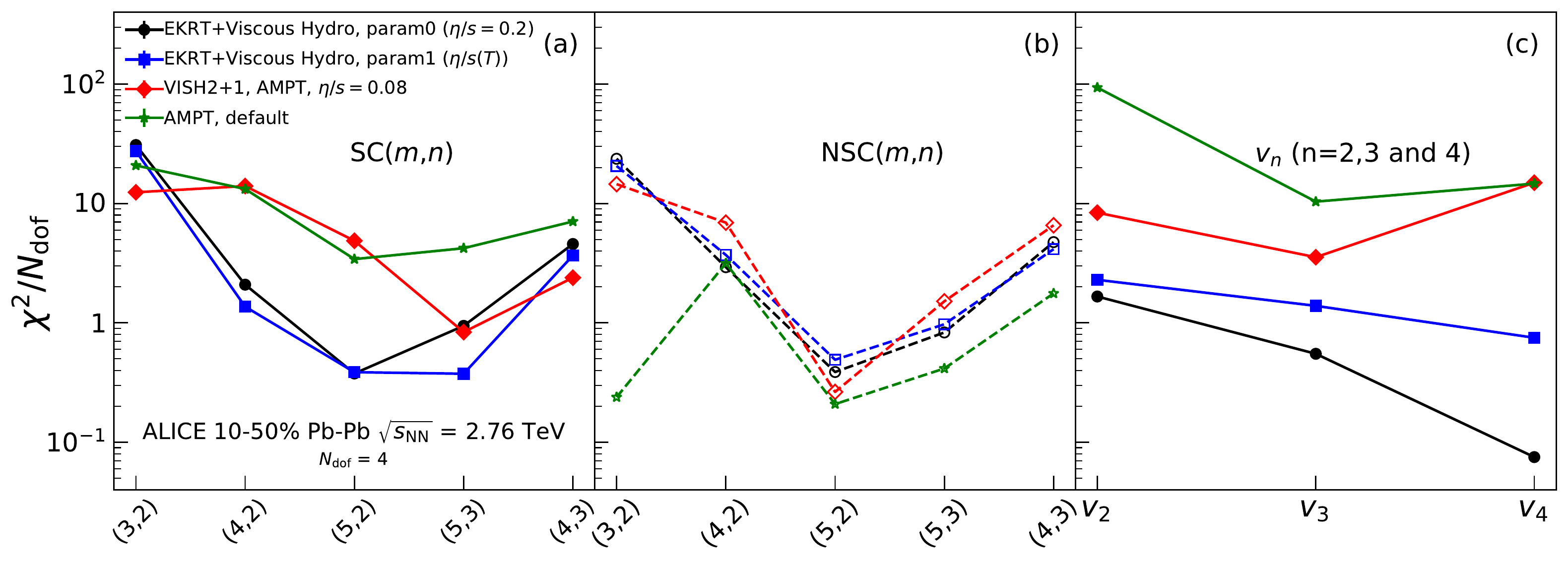}}
        \caption{The $\chisqndf$ values calculated by Eq.~(\ref{Eq:chisq}) are shown for SC($m$,$n$) (a), NSC($m$,$n$) (b), and individual harmonics $v_n$ (c). Results are for model calculations which are best in describing the SC observables for each of the three different types of models.}
        \label{fig:Figure_7}
              \end{center}
\end{figure}

The agreement of various model calculations with the data is quantified by calculating the $\chisqndf$,

\begin{equation}
\chi^2/N_{\rm dof} = \frac{1}{N_{\rm dof}} \sum_{i=1}^{N_{\rm dof}} \frac{{(y_i-f_i)}^{2}}{{\sigma_i}^{2}},
\label{Eq:chisq}
\end{equation}

where $y_i$ ($f_i$) is a measurement (model) value in a centrality bin $i$. The systematic and statistical errors from the data are combined in quadrature $\sigma_i = \sqrt{\sigma_{i, stat}^2 + \sigma_{i, syst}^2 + \sigma_{f_i, stat}^2}$ together with the statistical errors of the model calculations.
The total number of data samples $N_{\rm dof}$ in Eq.~(\ref{Eq:chisq}) is 4, which corresponds to the number of bins in the centrality range 10--50\% used in $\chisqndf$ calculations.
The $\chisqndf$ for model calculations which are best in describing the SC observables for each of the three different types of models are shown in Fig.~\ref{fig:Figure_7}.

The results for SC($m$,$n$) and NSC($m$,$n$) are presented in Figs.~\ref{fig:Figure_7}(a) and \ref{fig:Figure_7}(b), respectively. The $\chisqndf$ values for the individual flow harmonics $v_n$ for $n$ = 2--4 are shown in Fig.~\ref{fig:Figure_7}(c). We found that in the case of the calculations from VISH2+1 with AMPT initial conditions ($\eta/s$ = 0.08) and the default configuration of the AMPT model, the $\chisqndf$ values for SC($m$,$n$) are larger than those for NSC($m$,$n$). This reflects the fact that the individual flow harmonics $v_n$ are not well described by those models compared to event-by-event EKRT+viscous hydrodynamics. This is quantified in Fig.~\ref{fig:Figure_7}(c), where the $\chisqndf$ values for $v_n$ are much larger both for VISH2+1 and default AMPT calculations than event-by-event EKRT+viscous hydrodynamics.
The default configuration of the AMPT model gives the best $\chisqndf$ values for NSC($m$,$n$), especially for NSC(3,2). However, the $\chisqndf$ values of this model are largest for $v_n$ among the models especially for $v_2$.

The $\chisqndf$ values for $v_2$ and $v_3$ are significantly smaller than those for SC(3,2) and NSC(3,2) for all the hydrodynamic calculations. The $\chisqndf$ values for SC(4,2) and NSC(4,2) from event-by-event EKRT+viscous hydrodynamics are comparable to that for $v_2$ but larger than for $v_4$. The $\chisqndf$ for calculations for $v_n$ with constant $\eta/s$ = 0.20 (``param0'') are smaller than those with temperature-dependent $\eta/s$ parametrization with a minimal value of $\eta/s$ = 0.12 at the temperature around 150 MeV (``param1''), while an opposite trend is observed for SC($m$,$n$), in particular for SC(4,2) and SC(5,3). 
This illustrates that a combination of the SC($m$,$n$) observables with the individual flow harmonics $v_n$ may provide sensitivity to the temperature dependence of the $\eta/s(T)$ and together they allow for better constraints of the model parameters.

Even though the calculations from event-by-event EKRT+viscous hydrodynamics give the best $\chisqndf$ values for both SC($m$,$n$) and NSC($m$,$n$), the $\chisqndf$
values are large, especially for the observables which include $v_3$. 
Even with the best model calculations, the $\chisqndf$ value varies a lot depending on the model parameters and/or different order SC observables, which implies that the different order harmonic correlations have different sensitivity to the initial conditions and the system properties.

\subsection{Transverse momentum dependence of correlations between $v_2$ and $v_3$ and between $v_2$ and $v_4$}
\label{sec:ptdepsc}
\begin{figure}[t!]
             \begin{center}
              \resizebox{0.95\textwidth}{!}{\includegraphics{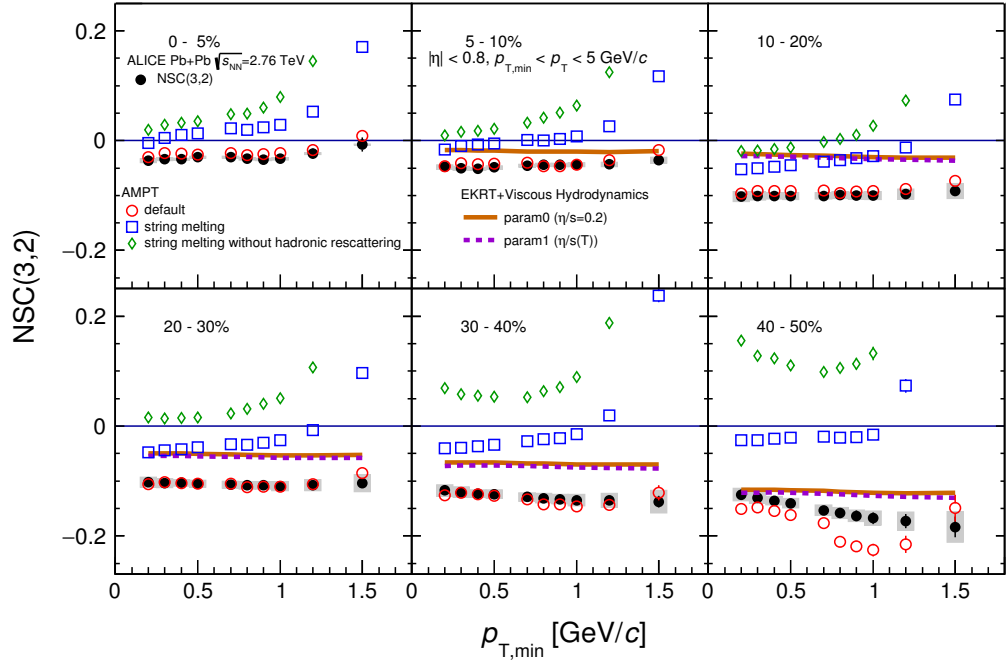}}
              \end{center}
             \caption{NSC(3,2) as a function of the minimum $p_{\rm T}$ cut in $\PbPb$ collisions at $\snn=2.76$~TeV. Results are compared to various AMPT configurations and event-by-event EKRT+viscous hydrodynamic calculations~\cite{Niemi:2015qia}.}
             \label{fig:Figure_8}
\end{figure}

\begin{figure}[t!]
             \begin{center}
              \resizebox{0.95\textwidth}{!}{\includegraphics{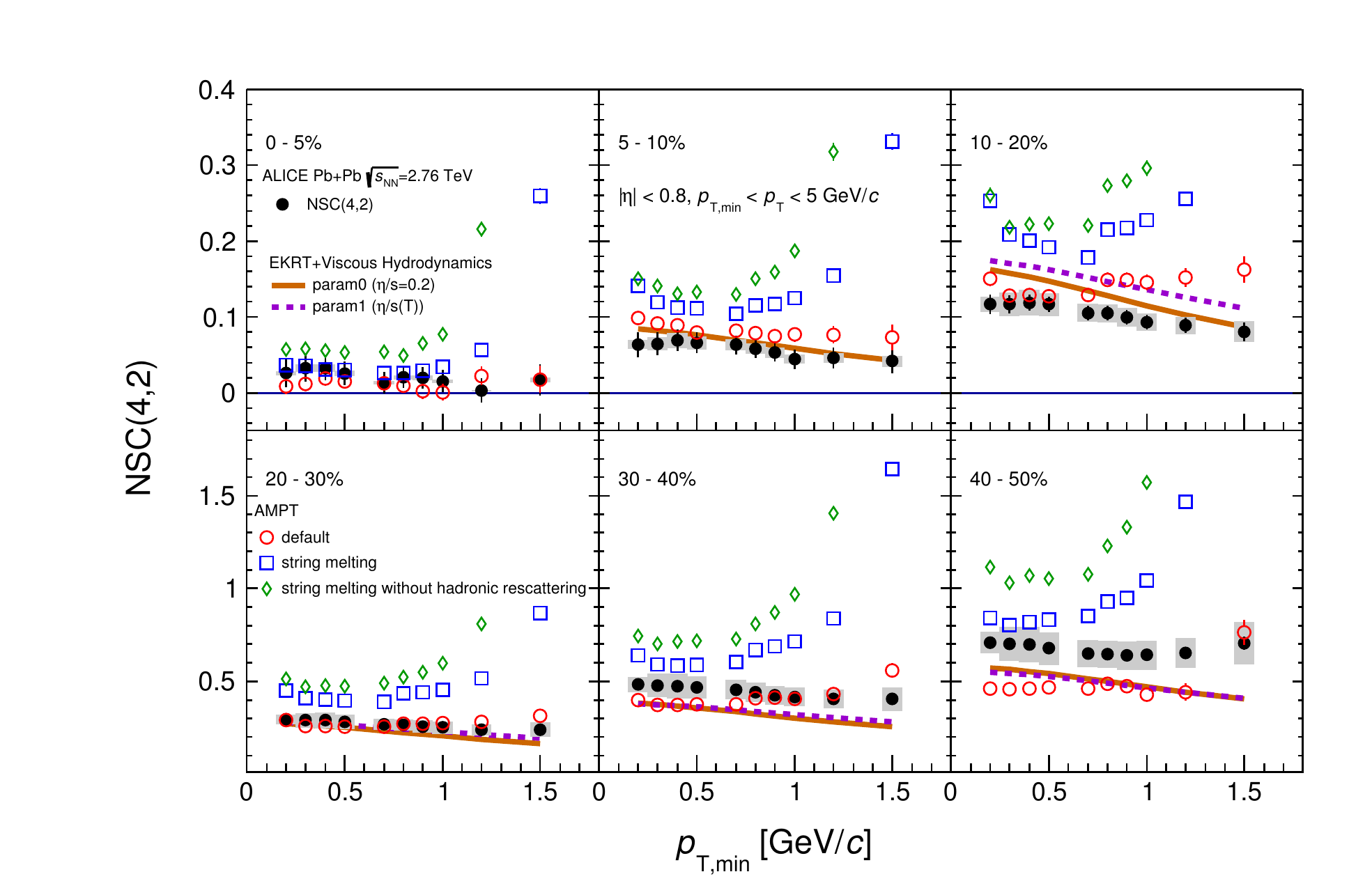}}
              \end{center}
             \caption{NSC(4,2) as a function of the minimum $p_{\rm T}$ cut in $\PbPb$ collisions at $\snn=2.76$~TeV. Results are compared to various AMPT configurations and event-by-event EKRT+viscous hydrodynamic calculations~\cite{Niemi:2015qia}.}
             \label{fig:Figure_9}
\end{figure}

The NSC(3,2) and NSC(4,2) observables as a function of $p_{\rm T, min}$ are compared to the {AMPT} simulations in Figs.~\ref{fig:Figure_8} and \ref{fig:Figure_9}, respectively.
The observed $p_{\rm T}$ dependence for NSC(3,2) in midcentral collisions is also seen in AMPT simulations for higher $p_{\rm T, min}$.
The default configuration of the AMPT reproduces NSC(3,2), while the other AMPT configurations predict a very strong $p_{\rm T}$ dependence above 1~GeV/$c$ and cannot describe the magnitudes of both NSC(3,2) and NSC(4,2) simultaneously.
In the case of NSC(3,2), the default AMPT model describes the magnitude and $p_{\rm T}$ dependence well in all collision centralities except for 40--50\%, where the model underestimates the data and shows a stronger $p_{\rm T}$ dependence than the data.
As for  NSC(4,2), the default AMPT configuration which describes NSC(3,2) can also reproduce the data well except for the 10--20\% and 40--50\% centralities.
Comparison of the string melting AMPT configuration with and without hadronic rescattering suggests that a very strong $p_{\rm T}$ dependence as well as the correlation strength are weakened by the hadronic rescattering.
Consequently, the observed weak $p_{\rm T}$ dependence may be due to hadronic rescattering. The relative contributions to the final-state particle distributions from partonic and hadronic stages need further study.

The event-by-event EKRT+viscous hydrodynamic calculations are also compared to the data in Figs.~\ref{fig:Figure_8} and \ref{fig:Figure_9}.
In the case of NSC(3,2), the hydrodynamic calculations underestimate the magnitude of the data as discussed in Sec.~\ref{sec:theory_allorder} and show very weak $p_{\rm T}$ dependence for all centralities.
The $p_{\rm T}$ dependence of NSC(3,2) is well captured by the model calculations in all collision centralities except for 40--50\%, where the data show stronger $p_{\rm T}$ dependence than the models. The difference between the model calculations with the two different parametrizations of $\eta/s(T)$ is very small. 
As for NSC(4,2), the model calculations overestimate the magnitude of the data in the 5--20\% centrality range and underestimate it in the centrality range 30--50\%. However, the $p_{\rm T}$ dependence is well described by the model calculations in all centrality ranges, while the difference of the model results for the two parametrizations in most centralities is rather small.

The observed moderate $p_{\rm T}$ dependence in midcentral collisions both for NSC(3,2) and NSC(4,2) might be an indication of possible viscous corrections to the equilibrium distribution at hadronic freeze-out, as predicted in Ref.~\cite{Niemi:2012aj}.
The comparisons to hydrodynamic models can further help us to understand the viscous corrections to the momentum distributions at hadronic freeze-out~\cite{Dusling:2009df,Luzum:2010ad,Teaney:2012ke,Molnar:2014fva,Niemi:2015qia}.
 
\section{Summary}
\label{sec:summary}
In this article, we report the centrality dependence of correlations between the higher order harmonics ($v_4$, $v_5$) and the lower order harmonics ($v_2$, $v_3$) as well as the transverse momentum dependence of the correlations between $v_3$ and $v_2$ and between $v_4$ and $v_2$.
The results are presented in terms of the symmetric cumulants SC($m$,$n$). It was demonstrated earlier in Ref.~\cite{ALICE:2016kpq} that SC($m$,$n$) is insensitive to nonflow effects and independent of symmetry plane correlations.

We have found that fluctuations of SC(3,2) and SC(4,3) are anticorrelated in all centralities while fluctuations of SC(4,2), SC(5,2), and SC(5,3) are correlated for all centralities. 
These measurements were compared to various hydrodynamic model calculations with different initial conditions as well as different parametrizations of the temperature dependence of $\eta/s$.
It is found that the different order harmonic correlations have different sensitivities to the initial conditions and the system properties. Therefore, they have discriminating power in separating the effects of $\eta/s$  from the initial conditions on the final-state particle anisotropies.
The comparisons to VISH2+1 calculations show that all the models with large $\eta/s$, regardless of the initial conditions, fail to describe the centrality dependence of higher order correlations. 
Based on the tested model parameters, the data favor small $\eta/s$ and the AMPT initial conditions. 

A quite clear separation of the correlation strength for different initial conditions is observed for these higher order harmonic correlations compared to the lower order.
The default configuration of the AMPT model describes well the normalized symmetric cumulants [NSC($m$,$n$)] for most centralities and for most combinations of harmonics which were considered. 
Finally, we have found that $v_3$ and $v_2$ as well as $v_4$ and $v_2$ correlations have moderate $p_{\rm T}$ dependence in midcentral collisions. This might be an indication of possible viscous corrections to the equilibrium distribution at hadronic freeze-out.
Together with the measurements of individual harmonics, the new results for SC($m$,$n$) and NSC($m$,$n$) can be used to further optimize model parameters and put better constraints on the initial conditions and the transport properties of nuclear matter in ultrarelativistic heavy-ion collisions.

               %%%%%%%%%%% put the body of the article here

%%%%%%%% acknowledgements
\newenvironment{acknowledgement}{\relax}{\relax}
\begin{acknowledgement}
\section*{Acknowledgements}
% Version: 2017-08-29

The ALICE Collaboration would like to thank all its engineers and technicians for their invaluable contributions to the construction of the experiment and the CERN accelerator teams for the outstanding performance of the LHC complex.
The ALICE Collaboration gratefully acknowledges the resources and support provided by all grid centers and the Worldwide LHC Computing Grid (WLCG) collaboration.
The ALICE Collaboration acknowledges the following funding agencies for their support in building and running the ALICE detector:
A. I. Alikhanyan National Science Laboratory (Yerevan Physics Institute) Foundation (ANSL), State Committee of Science and World Federation of Scientists (WFS), Armenia;
Austrian Academy of Sciences and Nationalstiftung f\"{u}r Forschung, Technologie und Entwicklung, Austria;
Ministry of Communications and High Technologies, National Nuclear Research Center, Azerbaijan;
Conselho Nacional de Desenvolvimento Cient\'{\i}fico e Tecnol\'{o}gico (CNPq), Universidade Federal do Rio Grande do Sul (UFRGS), Financiadora de Estudos e Projetos (Finep) and Funda\c{c}\~{a}o de Amparo \`{a} Pesquisa do Estado de S\~{a}o Paulo (FAPESP), Brazil;
Ministry of Science and Technology of China (MSTC), National Natural Science Foundation of China (NSFC), and Ministry of Education of China (MOEC) , China;
Ministry of Science, Education, and Sport and Croatian Science Foundation, Croatia;
Ministry of Education, Youth, and Sports of the Czech Republic, Czech Republic;
the Danish Council for Independent Research--Natural Sciences, the Carlsberg Foundation and Danish National Research Foundation (DNRF), Denmark;
Helsinki Institute of Physics (HIP), Finland;
Commissariat \`{a} l'Energie Atomique (CEA), Institut National de Physique Nucl\'{e}aire et de Physique des Particules (IN2P3), and Centre National de la Recherche Scientifique (CNRS), France;
Bundesministerium f\"{u}r Bildung, Wissenschaft, Forschung und Technologie (BMBF) and GSI Helmholtzzentrum f\"{u}r Schwerionenforschung GmbH, Germany;
General Secretariat for Research and Technology, Ministry of Education, Research and Religions, Greece;
National Research, Development and Innovation Office, Hungary;
Department of Atomic Energy Government of India (DAE) and Council of Scientific and Industrial Research (CSIR), New Delhi, India;
Indonesian Institute of Science, Indonesia;
Centro Fermi - Museo Storico della Fisica e Centro Studi e Ricerche Enrico Fermi and Istituto Nazionale di Fisica Nucleare (INFN), Italy;
Institute for Innovative Science and Technology , Nagasaki Institute of Applied Science (IIST), Japan Society for the Promotion of Science (JSPS) KAKENHI and Japanese Ministry of Education, Culture, Sports, Science and Technology (MEXT), Japan;
Consejo Nacional de Ciencia (CONACYT) y Tecnolog\'{i}a, through Fondo de Cooperaci\'{o}n Internacional en Ciencia y Tecnolog\'{i}a (FONCICYT) and Direcci\'{o}n General de Asuntos del Personal Academico (DGAPA), Mexico;
Nederlandse Organisatie voor Wetenschappelijk Onderzoek (NWO), Netherlands;
the Research Council of Norway, Norway;
Commission on Science and Technology for Sustainable Development in the South (COMSATS), Pakistan;
Pontificia Universidad Cat\'{o}lica del Per\'{u}, Peru;
Ministry of Science and Higher Education and National Science Centre, Poland;
Korea Institute of Science and Technology Information and National Research Foundation of Korea (NRF), Republic of Korea;
Ministry of Education and Scientific Research, Institute of Atomic Physics and Romanian National Agency for Science, Technology and Innovation, Romania;
Joint Institute for Nuclear Research (JINR), Ministry of Education and Science of the Russian Federation and National Research Centre Kurchatov Institute, Russia;
Ministry of Education, Science, Research and Sport of the Slovak Republic, Slovakia;
National Research Foundation of South Africa, South Africa;
Centro de Aplicaciones Tecnol\'{o}gicas y Desarrollo Nuclear (CEADEN), Cubaenerg\'{\i}a, Cuba, Ministerio de Ciencia e Innovacion and Centro de Investigaciones Energ\'{e}ticas, Medioambientales y Tecnol\'{o}gicas (CIEMAT), Spain;
Swedish Research Council (VR) and Knut and Alice Wallenberg Foundation (KAW), Sweden;
European Organization for Nuclear Research, Switzerland;
National Science and Technology Development Agency (NSDTA), Suranaree University of Technology (SUT) and Office of the Higher Education Commission under NRU project of Thailand, Thailand;
Turkish Atomic Energy Agency (TAEK), Turkey;
National Academy of  Sciences of Ukraine, Ukraine;
Science and Technology Facilities Council (STFC), United Kingdom;
and National Science Foundation of the United States of America (NSF) and United States Department of Energy, Office of Nuclear Physics (DOE NP), United States of America.    %%%%%%% get the lates version before submitting
\end{acknowledgement}

\bibliographystyle{utphys}
\bibliography{paper.bib}
%\newpage
%
%\input{}               %%%%%%%%%%% put your appendices here
% !TEX root = paper.tex
\appendix
\section{Model Comparisons of the Individual Flow Harmonics $v_n$}
\label{sec:vn}
As discussed in Sec.~\ref{sec:method}, NSC$(m,n)$ is expected to be insensitive to the magnitudes of $v_{m}$ and $v_{n}$ but SC$(m,n)$ has contributions from both the correlations between the two different flow harmonics and the individual harmonics $v_{n}$. Therefore, it is important to check how well the theoretical models used in Sec.~\ref{sec:theory} describe the measured $v_n$ data shown in Sec.~\ref{sec:results}. $v_n$ results presented in this section are for charged particles in the pseudorapidity range $|\eta| < 0.8$ and the transverse momentum range $0.2 < p_{\rm T} < 5.0$ GeV/$c$ as a function of collision centrality~\cite{Adam:2016izf}.

\begin{figure*}[h]
            \begin{center}
                       \resizebox{0.34\textwidth}{!}{\includegraphics{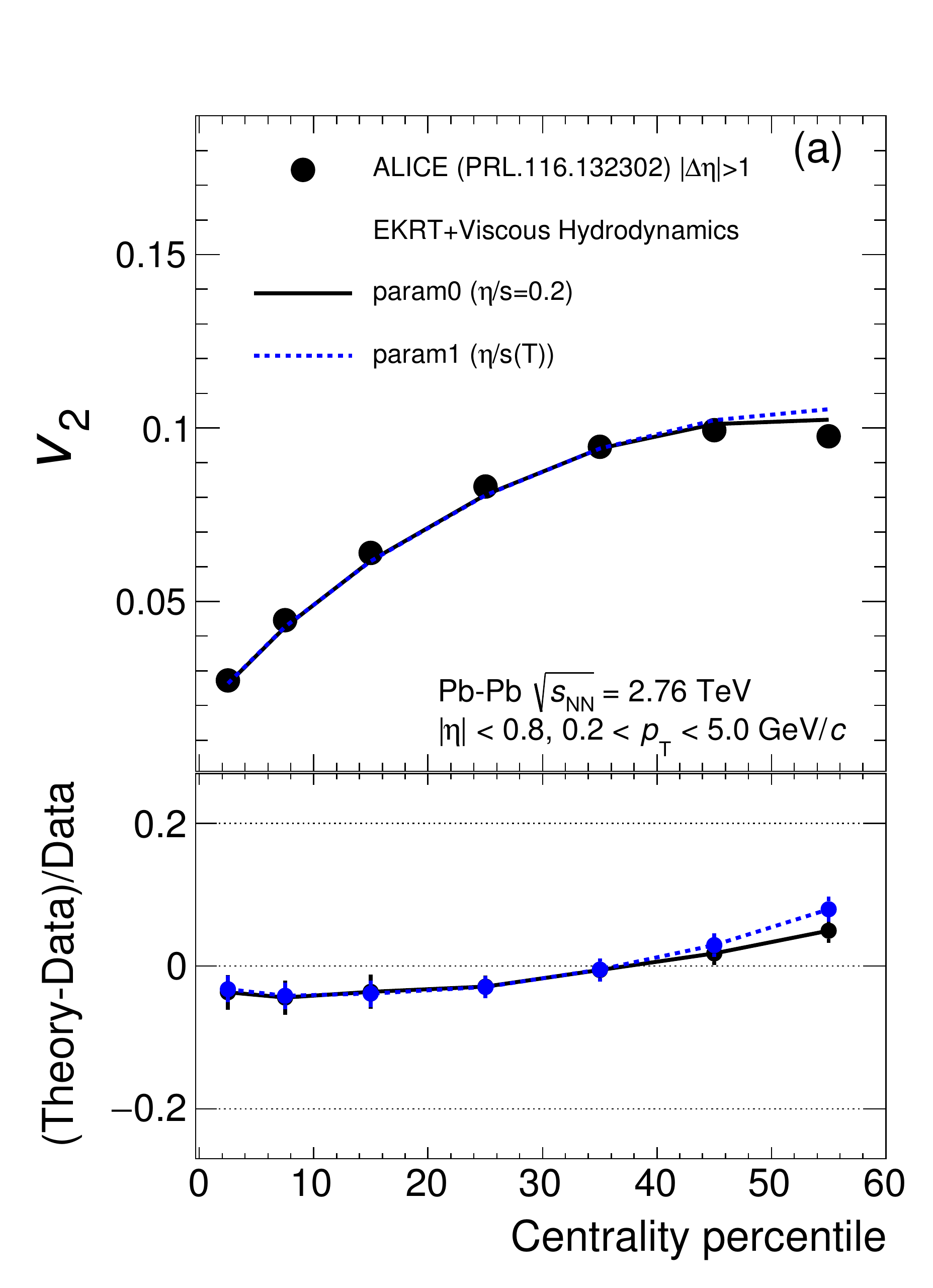}}\hspace{-0.27cm}
                       \resizebox{0.34\textwidth}{!}{\includegraphics{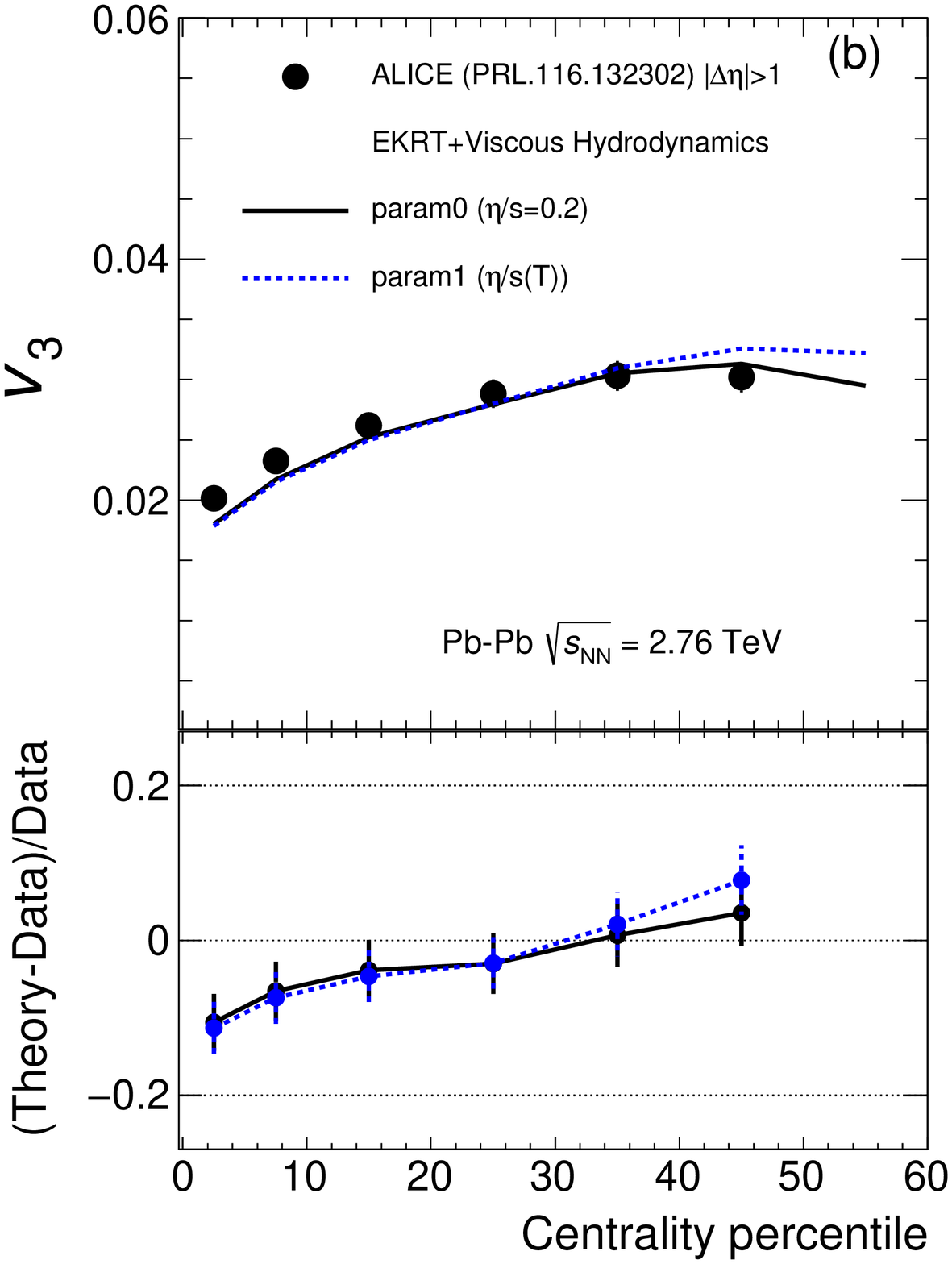}}\hspace{-0.27cm}
                       \resizebox{0.34\textwidth}{!}{\includegraphics{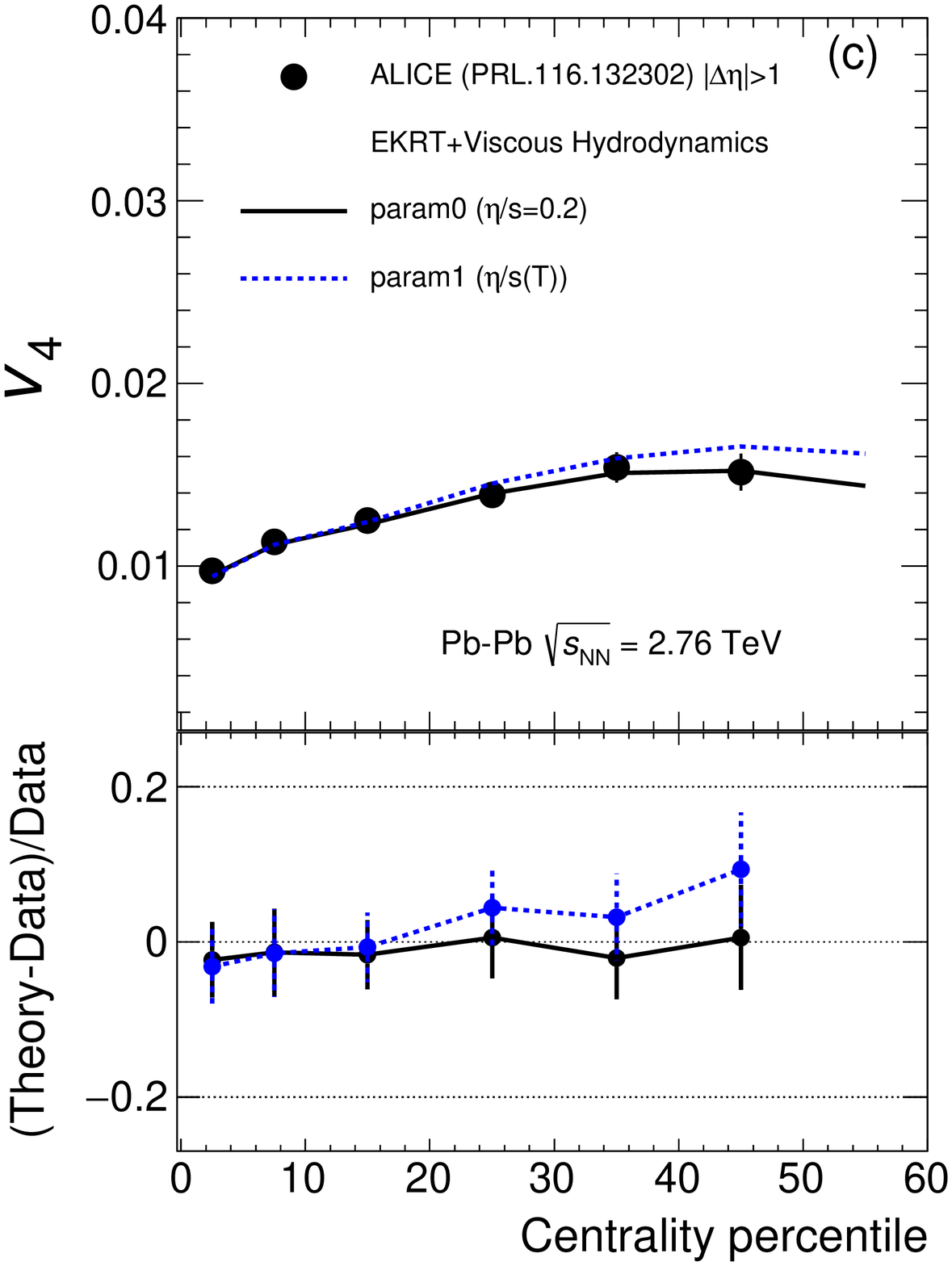}}
        \caption{The individual flow harmonics $v_n$ for $n$ = 2--4 in $\PbPb$ collisions at $\snn=2.76$~TeV~\cite{Adam:2016izf}. Results are compared to the event-by-event EKRT+viscous hydrodynamic calculations~\cite{Niemi:2015qia} for two different $\eta/s(T)$ parametrizations, labeled in the same way as in Ref.~\cite{Niemi:2015qia}.}
        \label{fig:Figure_A1}
              \end{center}
\end{figure*}

The measured $v_n$ for $n$ = 2--4 in $\PbPb$ collisions at $\snn=2.76$~TeV are compared to the event-by-event EKRT+viscous hydrodynamic calculations~\cite{Niemi:2015qia} in Fig.~\ref{fig:Figure_A1}. In these calculations, the initial conditions and $\eta/s$ parametrizations are chosen to reproduce the LHC $v_n$ data.
The calculations capture the centrality dependence of $v_n$ in the central and midcentral collisions within 5\% for $v_2$ and 10\% for $v_3$ and $v_4$.

\begin{figure*}[h]
            \begin{center}
                       \resizebox{0.34\textwidth}{!}{\includegraphics{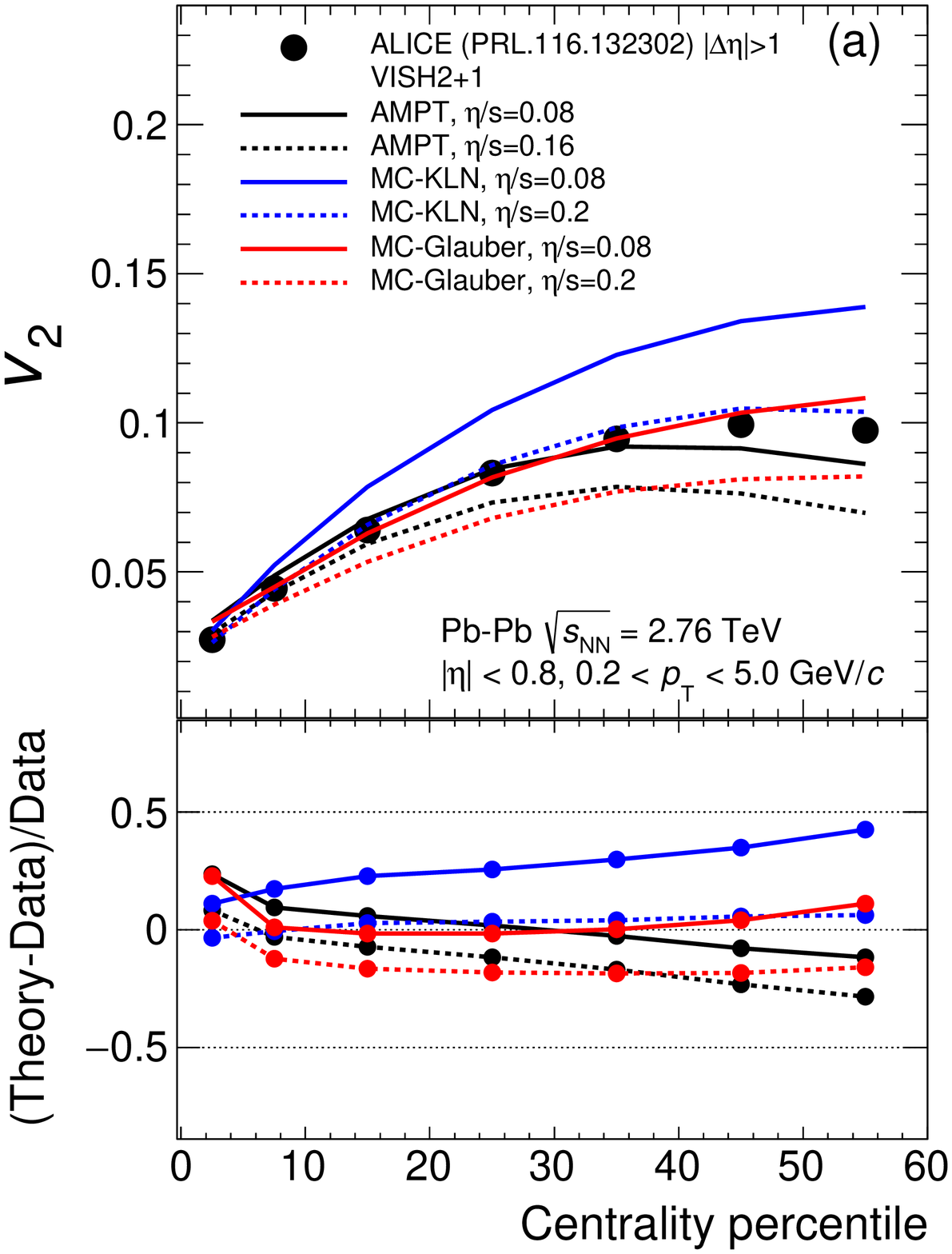}}\hspace{-0.27cm}
                       \resizebox{0.34\textwidth}{!}{\includegraphics{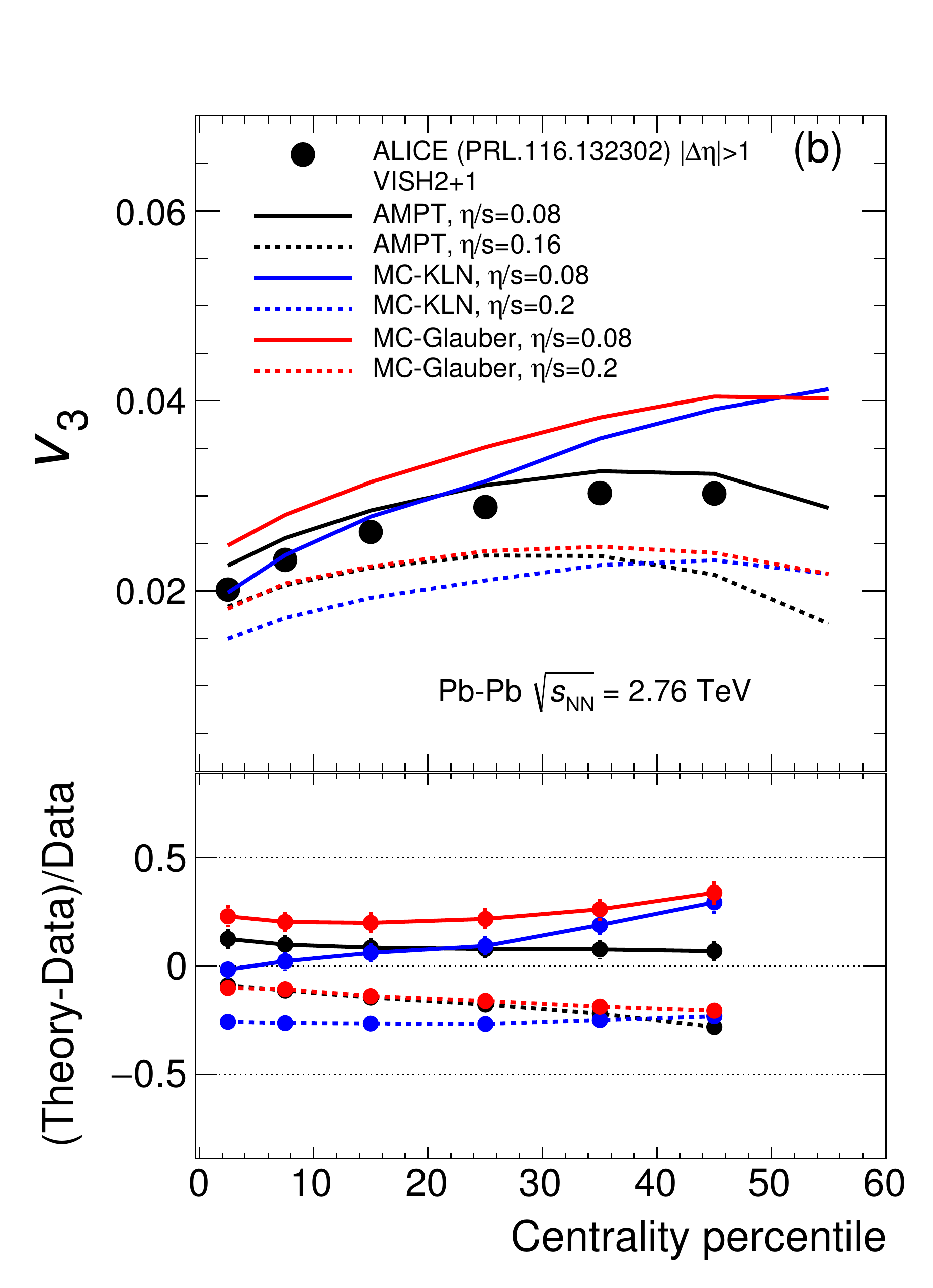}}\hspace{-0.27cm}
                       \resizebox{0.34\textwidth}{!}{\includegraphics{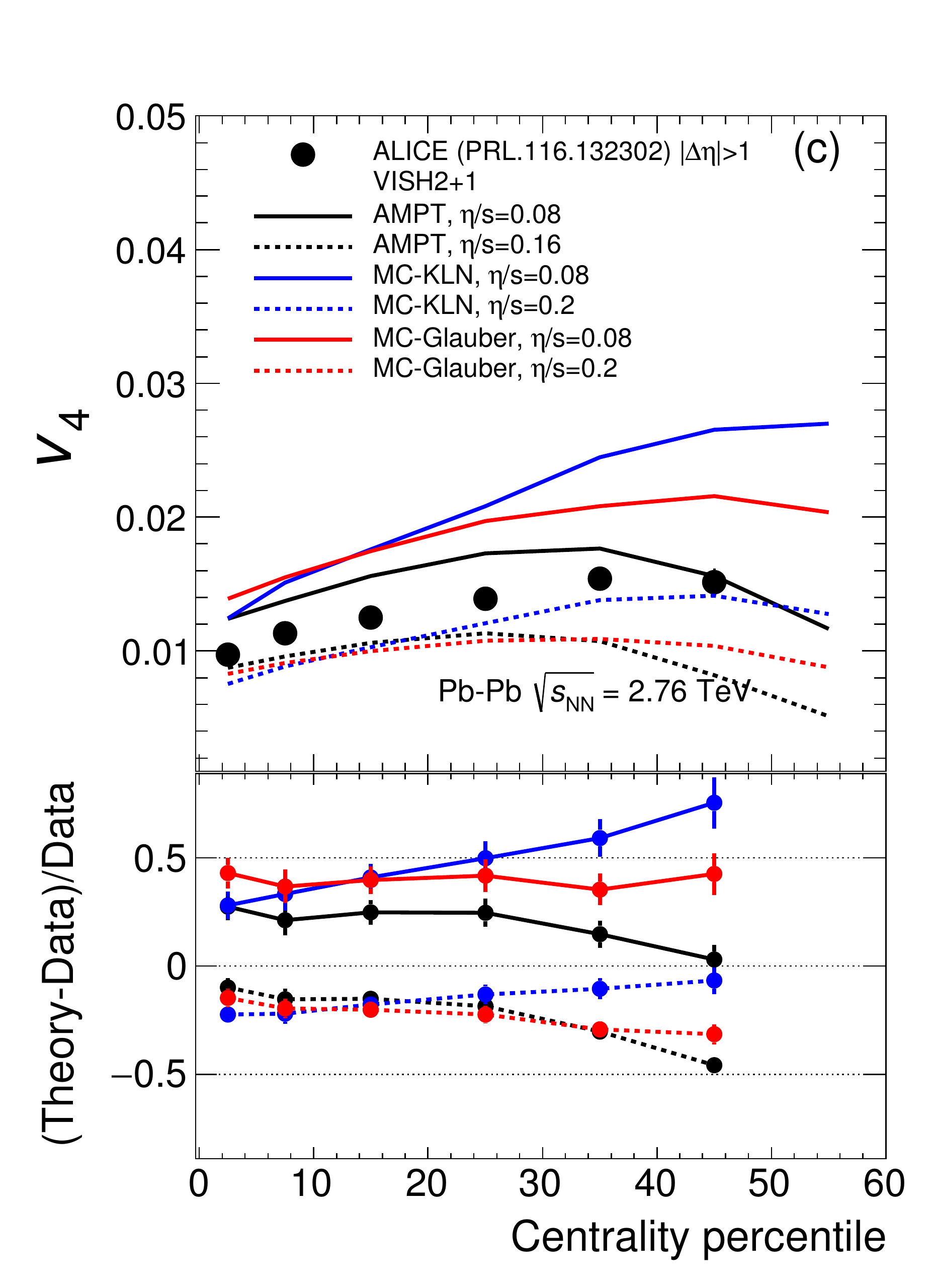}}
        \caption{The individual flow harmonics $v_n$ for $n$ = 2--4 in $\PbPb$ collisions at $\snn=2.76$~TeV~\cite{Adam:2016izf}. Results are compared to various VISH2+1 calculations~\cite{Zhu:2016puf}. Three initial conditions from AMPT, MC-KLN, and MC-Glauber are shown in different colors. The results for different $\eta/s$ values are shown as different line styles, the small shear viscosity ($\eta/s = 0.08$) are shown as solid lines, and large shear viscosities ($\eta/s = 0.2$ for MC-KLN and MC-Glauber and, 0.16 for AMPT) are drawn as dashed lines.}
        \label{fig:Figure_A2}
              \end{center}
\end{figure*}

The VISH2+1 calculations with various initial conditions and $\eta/s$ parameters are compared to the $v_n$ data in Fig.~\ref{fig:Figure_A2}.
Neither MC-Glauber nor MC-KLN initial conditions can simultaneously describe $v_2$, $v_3$, and $v_4$. In particular, for MC-Glauber initial conditions, VISH2+1 with $\eta/s$ = 0.08 can describe well $v_2$ from central to mid-central collisions, but overestimates $v_3$ and $v_4$ for the same centrality ranges. For MC-KLN initial conditions, VISH2+1 with $\eta/s$ = 0.20 reproduces $v_2$ but underestimates $v_3$ and $v_4$ for the presented centrality regions. The calculations with AMPT initial conditions improves the simultaneous descriptions of $v_n$ ($n$ = 2, 3, and 4). The overall difference to the data is quite large if all the model settings are considered, about 30\% for $v_n$ ($n$ = 2 and 3) and 50\% for $v_4$. The calculations with AMPT initial conditions reproduce the observed centrality dependence with an accuracy of 10--20\%.

\begin{figure*}[h]
            \begin{center}
                       \resizebox{0.34\textwidth}{!}{\includegraphics{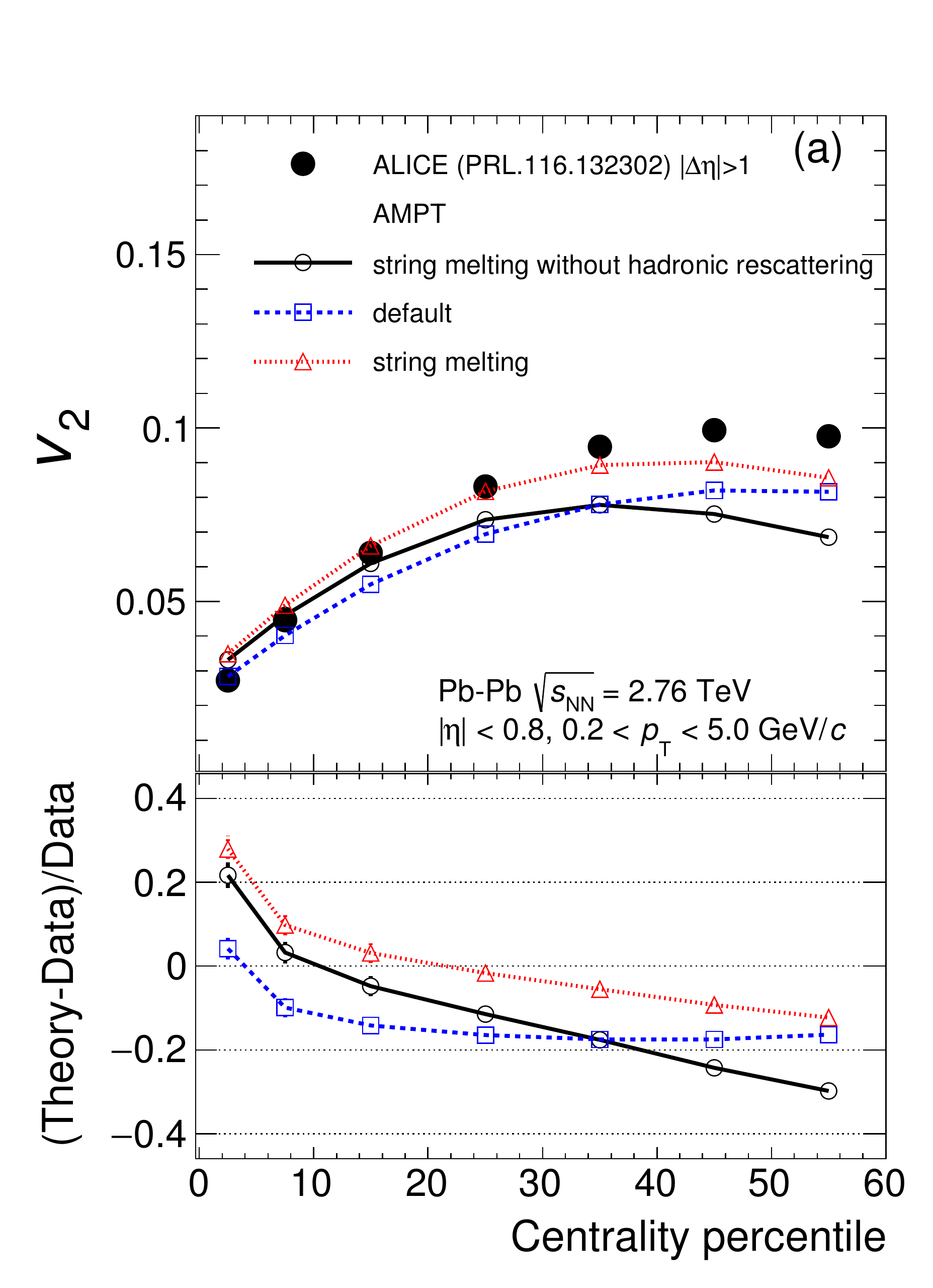}}\hspace{-0.27cm}
                       \resizebox{0.34\textwidth}{!}{\includegraphics{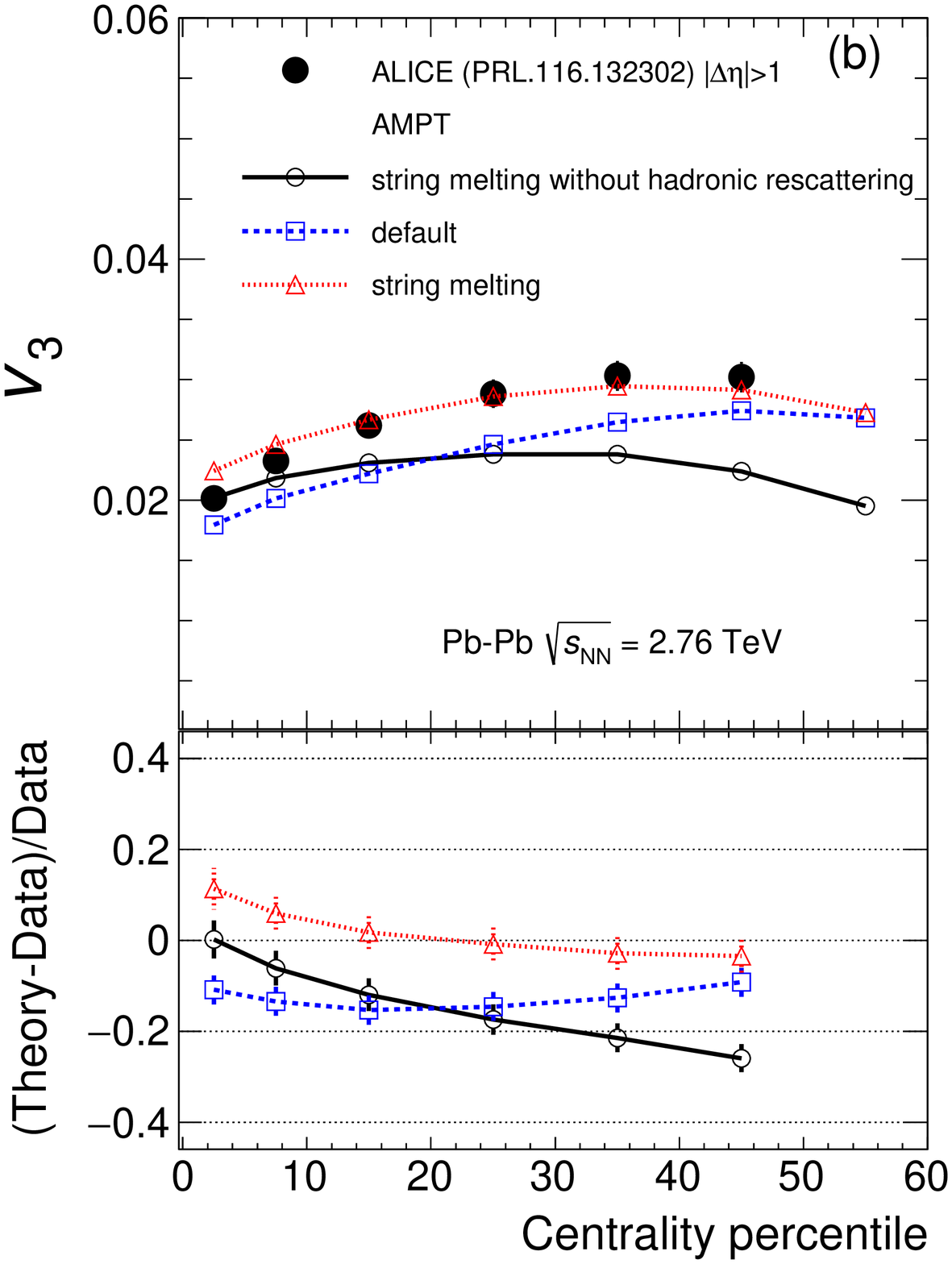}}\hspace{-0.27cm}
                       \resizebox{0.34\textwidth}{!}{\includegraphics{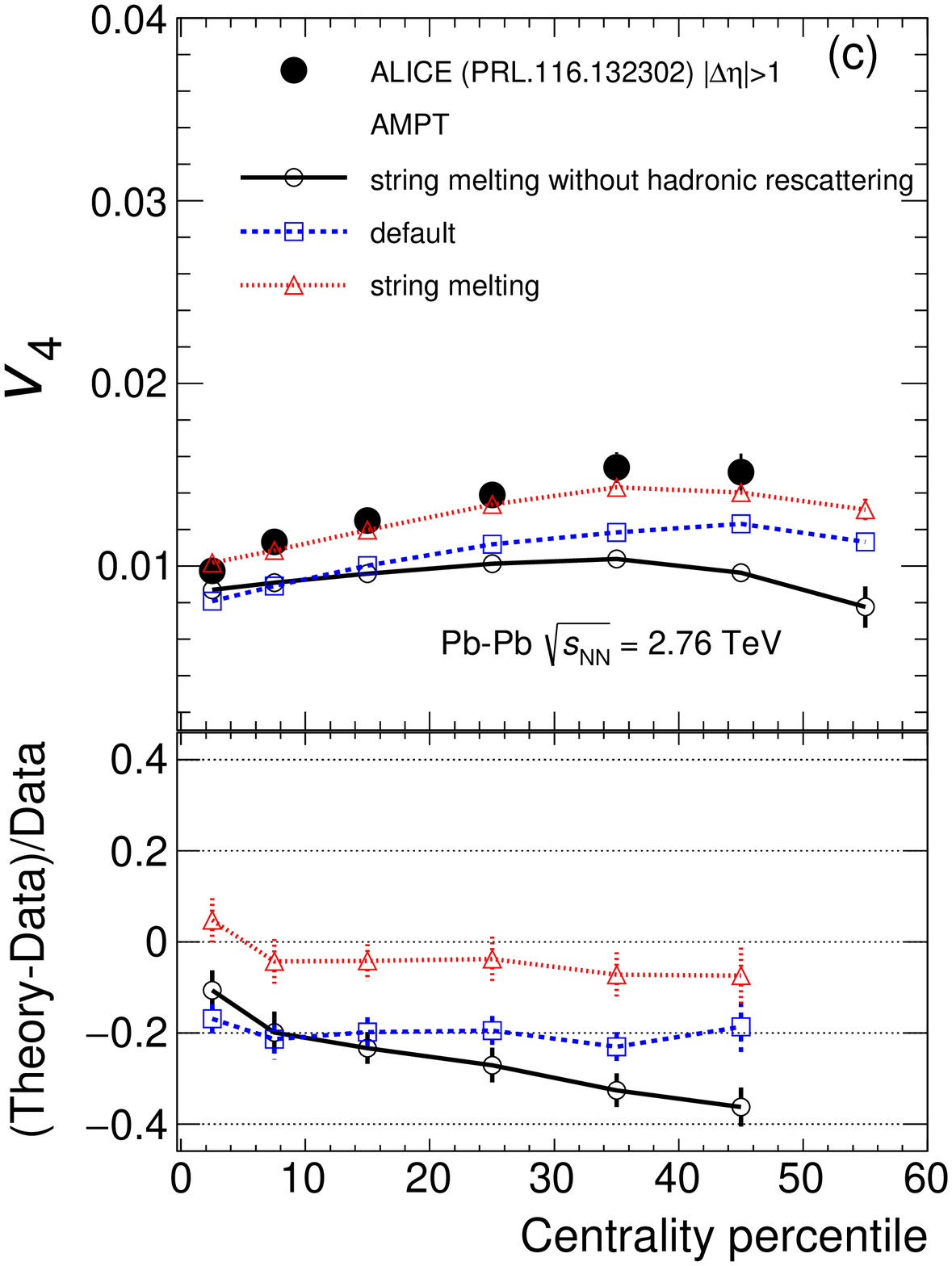}}
        \caption{The individual flow harmonics $v_n$ ($n$ = 2, 3, and 4) in $\PbPb$ collisions at $\snn=2.76$~TeV~\cite{Adam:2016izf}. Results are compared to various AMPT models.}
        \label{fig:Figure_A3}
              \end{center}
\end{figure*}

The AMPT calculations with various configurations are compared to the $v_n$ data in Fig.~\ref{fig:Figure_A3}.
The string melting version of AMPT~\cite{Lin:2001zk,Lin:2004en} reasonably reproduces $v_n$ as shown in Fig.~\ref{fig:Figure_A3} within 20\% for $v_2$ and 10\% for $v_3$ and $v_4$. The version based on the string melting configuration without the hadronic rescattering phase underestimates the data compared to the calculations with the string melting version of AMPT, which demonstrates that a large fraction of the flow is developed during the late hadronic rescattering stage in the string melting version of AMPT.
The default version of AMPT underestimates $v_n$ for $n$ = 2--4 by $\approx$ 20\%. It should be noted that the default AMPT model can describe the normalized symmetric cumulants [NSC($m$,$n$)] quantitively for most centralities while the string melting AMPT model fails to describe them.

\begin{figure*}[h]
            \begin{center}
                       \resizebox{0.34\textwidth}{!}{\includegraphics{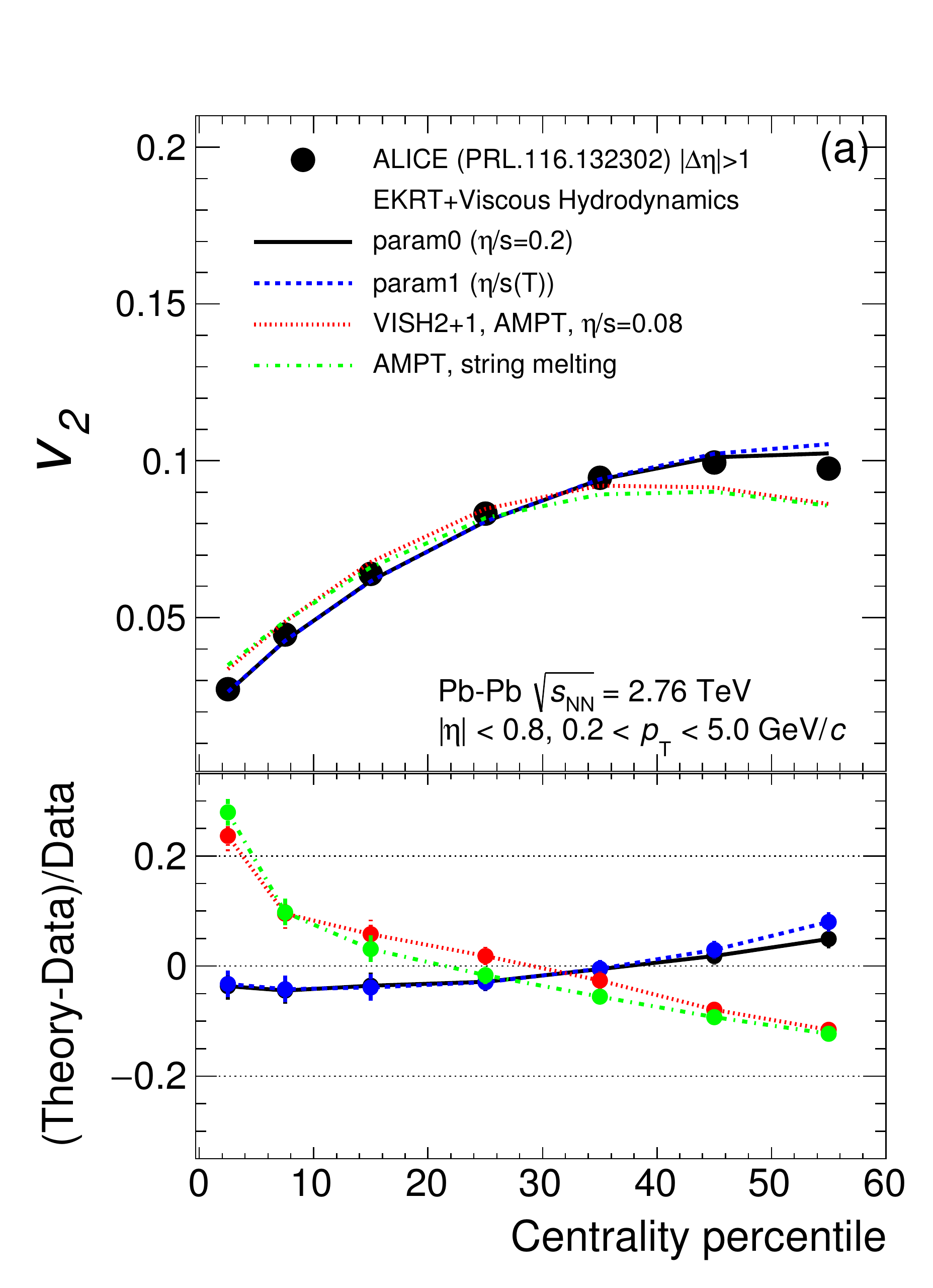}}\hspace{-0.27cm}
                       \resizebox{0.34\textwidth}{!}{\includegraphics{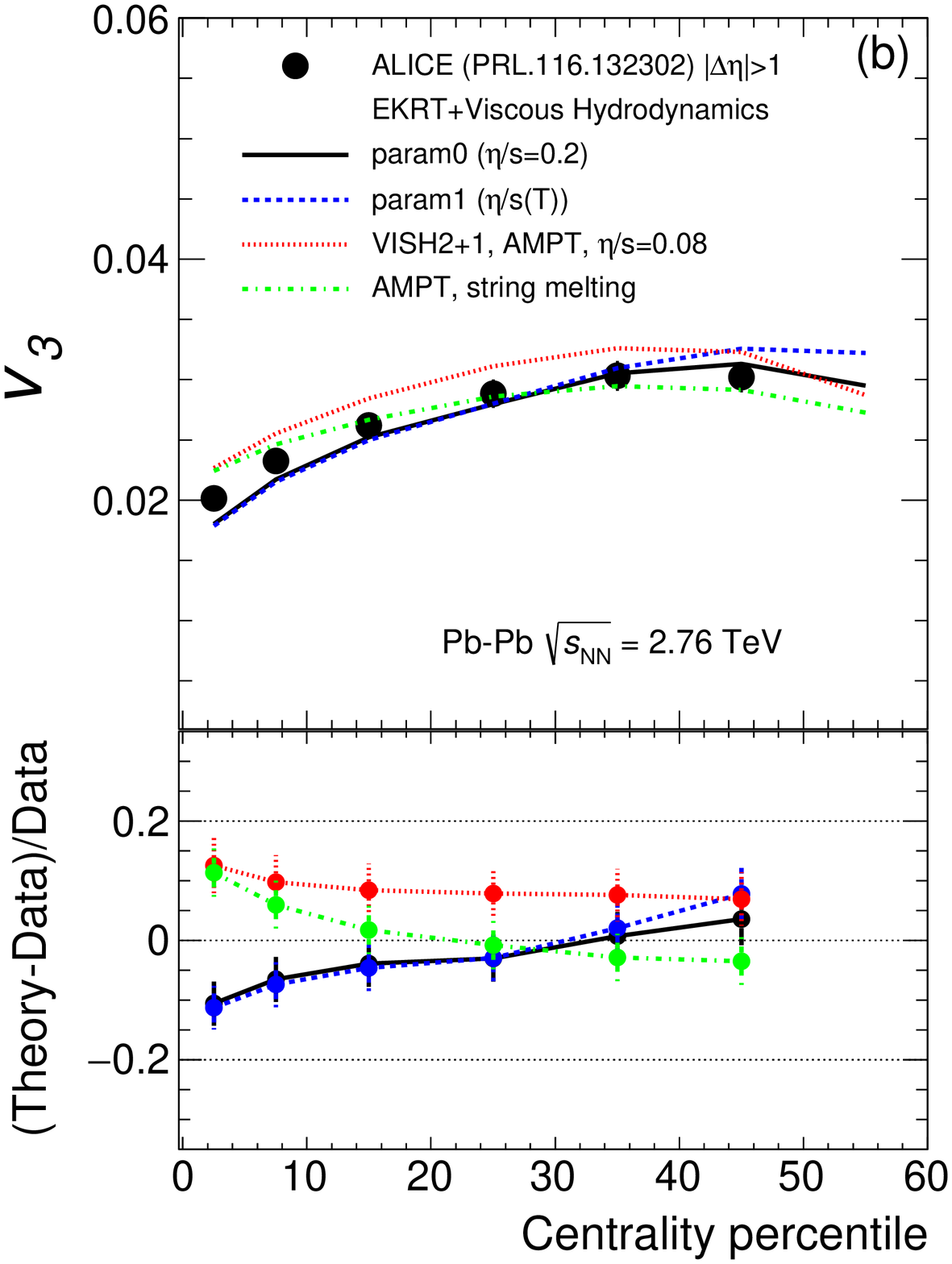}}\hspace{-0.27cm}
                       \resizebox{0.34\textwidth}{!}{\includegraphics{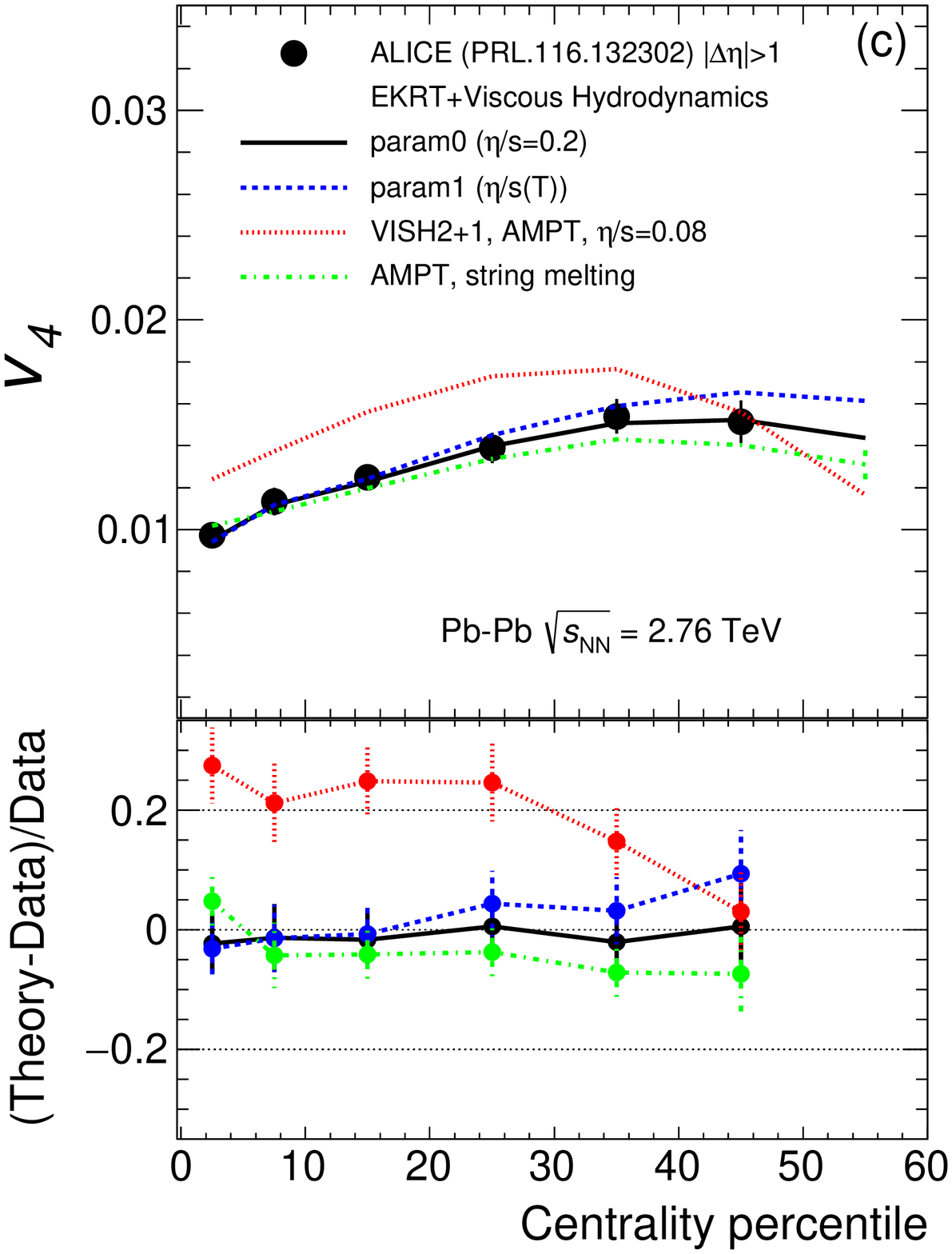}}
        \caption{The individual flow harmonics $v_n$ for $n$ = 2--4 in $\PbPb$ collisions at $\snn=2.76$~TeV~\cite{Adam:2016izf}. Results are compared with selected calculations from three different types of models which are best in describing $v_n$ coefficients.}
        \label{fig:Figure_A4}
              \end{center}
\end{figure*}

Finally, few selected calculations from three theoretical models which describe the $v_n$ data best are shown in Fig.~\ref{fig:Figure_A4}.
The calculations from event-by-event EKRT+viscous hydrodynamics, VISH2+1 with AMPT initial conditions ($\eta/s$ = 0.08) and the string melting version of AMPT give the best description of the individual flow harmonics $v_n$ ($n$ = 2, 3, and 4) with an accuracy of 5--20\%. The centrality dependence differs in the three models as well as in the different order flow harmonics.
Together with SC$(m,n)$ and NSC$(m,n)$, the simultaneous description of individual flow harmonics $v_n$ at all orders is necessary to further optimize model parameters and put better constraints on the initial conditions and the transport properties of nuclear matter in ultrarelativistic heavy-ion collisions.

\clearpage
\newpage

%%%%%%%%% appendix with author list
\section{The ALICE Collaboration}
\label{app:collab}
% Collaboration: CERN-LHC-ALICE
% Generation Date is 2017-Aug-21

% How to use:
%%%%%%%%% appendix with author list
%\appendix
%\section{The ALICE Collaboration}
%\label{app:collab}
%\input{Alice_Authorslist_XXXX-Axx-XX.tex}
\begingroup
\small
\begin{flushleft}
S.~Acharya\Irefn{org137}\And 
J.~Adam\Irefn{org96}\And 
D.~Adamov\'{a}\Irefn{org93}\And 
J.~Adolfsson\Irefn{org32}\And 
M.M.~Aggarwal\Irefn{org98}\And 
G.~Aglieri Rinella\Irefn{org33}\And 
M.~Agnello\Irefn{org29}\And 
N.~Agrawal\Irefn{org46}\And 
Z.~Ahammed\Irefn{org137}\And 
N.~Ahmad\Irefn{org15}
% ...
\And
\renewcommand\labelenumi{\textsuperscript{\theenumi}~}

\section*{Affiliation notes}
\renewcommand\theenumi{\roman{enumi}}
\begin{Authlist}
\item \Adef{org*}Deceased
\item \Adef{orgI}Dipartimento DET del Politecnico di Torino, Turin, Italy
\item \Adef{orgII}Georgia State University, Atlanta, Georgia, United States
\item \Adef{orgIII}M.V. Lomonosov Moscow State University, D.V. Skobeltsyn Institute of Nuclear, Physics, Moscow, Russia
\item \Adef{orgIV}Department of Applied Physics, Aligarh Muslim University, Aligarh, India
\item \Adef{orgV}Institute of Theoretical Physics, University of Wroclaw, Poland
\end{Authlist}

\section*{Collaboration Institutes}
\renewcommand\theenumi{\arabic{enumi}~}
\begin{Authlist}
\item \Idef{org1}A.I. Alikhanyan National Science Laboratory (Yerevan Physics Institute) Foundation, Yerevan, Armenia
\item \Idef{org2}Benem\'{e}rita Universidad Aut\'{o}noma de Puebla, Puebla, Mexico
\item \Idef{org3}Bogolyubov Institute for Theoretical Physics, Kiev, Ukraine
\item \Idef{org4}Bose Institute, Department of Physics  and Centre for Astroparticle Physics and Space Science (CAPSS), Kolkata, India
\item \Idef{org5}Budker Institute for Nuclear Physics, Novosibirsk, Russia
\item \Idef{org6}California Polytechnic State University, San Luis Obispo, California, United States
\item \Idef{org7}Central China Normal University, Wuhan, China
\item \Idef{org8}Centre de Calcul de l'IN2P3, Villeurbanne, Lyon, France
\item \Idef{org9}Centro de Aplicaciones Tecnol\'{o}gicas y Desarrollo Nuclear (CEADEN), Havana, Cuba
\item \Idef{org10}Centro de Investigaci\'{o}n y de Estudios Avanzados (CINVESTAV), Mexico City and M\'{e}rida, Mexico
\item \Idef{org11}Centro Fermi - Museo Storico della Fisica e Centro Studi e Ricerche ``Enrico Fermi', Rome, Italy
\item \Idef{org12}Chicago State University, Chicago, Illinois, United States
\item \Idef{org13}China Institute of Atomic Energy, Beijing, China
\item \Idef{org14}COMSATS Institute of Information Technology (CIIT), Islamabad, Pakistan
\item \Idef{org15}Department of Physics, Aligarh Muslim University, Aligarh, India
\item \Idef{org16}Department of Physics, Ohio State University, Columbus, Ohio, United States
\item \Idef{org17}Department of Physics, Pusan National University, Pusan, Republic of Korea
\item \Idef{org18}Department of Physics, Sejong University, Seoul, Republic of Korea
\item \Idef{org19}Department of Physics, University of Oslo, Oslo, Norway
\item \Idef{org20}Department of Physics and Technology, University of Bergen, Bergen, Norway
\item \Idef{org21}Dipartimento di Fisica dell'Universit\`{a} 'La Sapienza' and Sezione INFN, Rome, Italy
\item \Idef{org22}Dipartimento di Fisica dell'Universit\`{a} and Sezione INFN, Cagliari, Italy
\item \Idef{org23}Dipartimento di Fisica dell'Universit\`{a} and Sezione INFN, Trieste, Italy
\item \Idef{org24}Dipartimento di Fisica dell'Universit\`{a} and Sezione INFN, Turin, Italy
\item \Idef{org25}Dipartimento di Fisica e Astronomia dell'Universit\`{a} and Sezione INFN, Bologna, Italy
\item \Idef{org26}Dipartimento di Fisica e Astronomia dell'Universit\`{a} and Sezione INFN, Catania, Italy
\item \Idef{org27}Dipartimento di Fisica e Astronomia dell'Universit\`{a} and Sezione INFN, Padova, Italy
\item \Idef{org28}Dipartimento di Fisica `E.R.~Caianiello' dell'Universit\`{a} and Gruppo Collegato INFN, Salerno, Italy
\item \Idef{org29}Dipartimento DISAT del Politecnico and Sezione INFN, Turin, Italy
\item \Idef{org30}Dipartimento di Scienze e Innovazione Tecnologica dell'Universit\`{a} del Piemonte Orientale and INFN Sezione di Torino, Alessandria, Italy
\item \Idef{org31}Dipartimento Interateneo di Fisica `M.~Merlin' and Sezione INFN, Bari, Italy
\item \Idef{org32}Division of Experimental High Energy Physics, University of Lund, Lund, Sweden
\item \Idef{org33}European Organization for Nuclear Research (CERN), Geneva, Switzerland
\item \Idef{org34}Excellence Cluster Universe, Technische Universit\"{a}t M\"{u}nchen, Munich, Germany
\item \Idef{org35}Faculty of Engineering, Bergen University College, Bergen, Norway
\item \Idef{org36}Faculty of Mathematics, Physics and Informatics, Comenius University, Bratislava, Slovakia
\item \Idef{org37}Faculty of Nuclear Sciences and Physical Engineering, Czech Technical University in Prague, Prague, Czech Republic
\item \Idef{org38}Faculty of Science, P.J.~\v{S}af\'{a}rik University, Ko\v{s}ice, Slovakia
\item \Idef{org39}Faculty of Technology, Buskerud and Vestfold University College, Tonsberg, Norway
\item \Idef{org40}Frankfurt Institute for Advanced Studies, Johann Wolfgang Goethe-Universit\"{a}t Frankfurt, Frankfurt, Germany
\item \Idef{org41}Gangneung-Wonju National University, Gangneung, Republic of Korea
\item \Idef{org42}Gauhati University, Department of Physics, Guwahati, India
\item \Idef{org43}Helmholtz-Institut f\"{u}r Strahlen- und Kernphysik, Rheinische Friedrich-Wilhelms-Universit\"{a}t Bonn, Bonn, Germany
\item \Idef{org44}Helsinki Institute of Physics (HIP), Helsinki, Finland
\item \Idef{org45}Hiroshima University, Hiroshima, Japan
\item \Idef{org46}Indian Institute of Technology Bombay (IIT), Mumbai, India
\item \Idef{org47}Indian Institute of Technology Indore, Indore, India
\item \Idef{org48}Indonesian Institute of Sciences, Jakarta, Indonesia
\item \Idef{org49}INFN, Laboratori Nazionali di Frascati, Frascati, Italy
\item \Idef{org50}INFN, Laboratori Nazionali di Legnaro, Legnaro, Italy
\item \Idef{org51}INFN, Sezione di Bari, Bari, Italy
\item \Idef{org52}INFN, Sezione di Bologna, Bologna, Italy
\item \Idef{org53}INFN, Sezione di Cagliari, Cagliari, Italy
\item \Idef{org54}INFN, Sezione di Catania, Catania, Italy
\item \Idef{org55}INFN, Sezione di Padova, Padova, Italy
\item \Idef{org56}INFN, Sezione di Roma, Rome, Italy
\item \Idef{org57}INFN, Sezione di Torino, Turin, Italy
\item \Idef{org58}INFN, Sezione di Trieste, Trieste, Italy
\item \Idef{org59}Inha University, Incheon, Republic of Korea
\item \Idef{org60}Institut de Physique Nucl\'eaire d'Orsay (IPNO), Universit\'e Paris-Sud, CNRS-IN2P3, Orsay, France
\item \Idef{org61}Institute for Nuclear Research, Academy of Sciences, Moscow, Russia
\item \Idef{org62}Institute for Subatomic Physics of Utrecht University, Utrecht, Netherlands
\item \Idef{org63}Institute for Theoretical and Experimental Physics, Moscow, Russia
\item \Idef{org64}Institute of Experimental Physics, Slovak Academy of Sciences, Ko\v{s}ice, Slovakia
\item \Idef{org65}Institute of Physics, Academy of Sciences of the Czech Republic, Prague, Czech Republic
\item \Idef{org66}Institute of Physics, Bhubaneswar, India
\item \Idef{org67}Institute of Space Science (ISS), Bucharest, Romania
\item \Idef{org68}Institut f\"{u}r Informatik, Johann Wolfgang Goethe-Universit\"{a}t Frankfurt, Frankfurt, Germany
\item \Idef{org69}Institut f\"{u}r Kernphysik, Johann Wolfgang Goethe-Universit\"{a}t Frankfurt, Frankfurt, Germany
\item \Idef{org70}Institut f\"{u}r Kernphysik, Westf\"{a}lische Wilhelms-Universit\"{a}t M\"{u}nster, M\"{u}nster, Germany
\item \Idef{org71}Instituto de Ciencias Nucleares, Universidad Nacional Aut\'{o}noma de M\'{e}xico, Mexico City, Mexico
\item \Idef{org72}Instituto de F\'{i}sica, Universidade Federal do Rio Grande do Sul (UFRGS), Porto Alegre, Brazil
\item \Idef{org73}Instituto de F\'{\i}sica, Universidad Nacional Aut\'{o}noma de M\'{e}xico, Mexico City, Mexico
\item \Idef{org74}IRFU, CEA, Universit\'{e} Paris-Saclay, Saclay, France
\item \Idef{org75}iThemba LABS, National Research Foundation, Somerset West, South Africa
\item \Idef{org76}Joint Institute for Nuclear Research (JINR), Dubna, Russia
\item \Idef{org77}Konkuk University, Seoul, Republic of Korea
\item \Idef{org78}Korea Institute of Science and Technology Information, Daejeon, Republic of Korea
\item \Idef{org79}KTO Karatay University, Konya, Turkey
\item \Idef{org80}Laboratoire de Physique Subatomique et de Cosmologie, Universit\'{e} Grenoble-Alpes, CNRS-IN2P3, Grenoble, France
\item \Idef{org81}Lawrence Berkeley National Laboratory, Berkeley, California, United States
\item \Idef{org82}Moscow Engineering Physics Institute, Moscow, Russia
\item \Idef{org83}Nagasaki Institute of Applied Science, Nagasaki, Japan
\item \Idef{org84}National and Kapodistrian University of Athens, Physics Department, Athens, Greece
\item \Idef{org85}National Centre for Nuclear Studies, Warsaw, Poland
\item \Idef{org86}National Institute for Physics and Nuclear Engineering, Bucharest, Romania
\item \Idef{org87}National Institute of Science Education and Research, HBNI, Jatni, India
\item \Idef{org88}National Nuclear Research Center, Baku, Azerbaijan
\item \Idef{org89}National Research Centre Kurchatov Institute, Moscow, Russia
\item \Idef{org90}Niels Bohr Institute, University of Copenhagen, Copenhagen, Denmark
\item \Idef{org91}Nikhef, Nationaal instituut voor subatomaire fysica, Amsterdam, Netherlands
\item \Idef{org92}Nuclear Physics Group, STFC Daresbury Laboratory, Daresbury, United Kingdom
\item \Idef{org93}Nuclear Physics Institute, Academy of Sciences of the Czech Republic, \v{R}e\v{z} u Prahy, Czech Republic
\item \Idef{org94}Oak Ridge National Laboratory, Oak Ridge, Tennessee, United States
\item \Idef{org95}Petersburg Nuclear Physics Institute, Gatchina, Russia
\item \Idef{org96}Physics Department, Creighton University, Omaha, Nebraska, United States
\item \Idef{org97}Physics department, Faculty of science, University of Zagreb, Zagreb, Croatia
\item \Idef{org98}Physics Department, Panjab University, Chandigarh, India
\item \Idef{org99}Physics Department, University of Cape Town, Cape Town, South Africa
\item \Idef{org100}Physics Department, University of Jammu, Jammu, India
\item \Idef{org101}Physics Department, University of Rajasthan, Jaipur, India
\item \Idef{org102}Physikalisches Institut, Eberhard Karls Universit\"{a}t T\"{u}bingen, T\"{u}bingen, Germany
\item \Idef{org103}Physikalisches Institut, Ruprecht-Karls-Universit\"{a}t Heidelberg, Heidelberg, Germany
\item \Idef{org104}Physik Department, Technische Universit\"{a}t M\"{u}nchen, Munich, Germany
\item \Idef{org105}Purdue University, West Lafayette, Indiana, United States
\item \Idef{org106}Research Division and ExtreMe Matter Institute EMMI, GSI Helmholtzzentrum f\"ur Schwerionenforschung GmbH, Darmstadt, Germany
\item \Idef{org107}Rudjer Bo\v{s}kovi\'{c} Institute, Zagreb, Croatia
\item \Idef{org108}Russian Federal Nuclear Center (VNIIEF), Sarov, Russia
\item \Idef{org109}Saha Institute of Nuclear Physics, Kolkata, India
\item \Idef{org110}School of Physics and Astronomy, University of Birmingham, Birmingham, United Kingdom
\item \Idef{org111}Secci\'{o}n F\'{\i}sica, Departamento de Ciencias, Pontificia Universidad Cat\'{o}lica del Per\'{u}, Lima, Peru
\item \Idef{org112}SSC IHEP of NRC Kurchatov institute, Protvino, Russia
\item \Idef{org113}Stefan Meyer Institut f\"{u}r Subatomare Physik (SMI), Vienna, Austria
\item \Idef{org114}SUBATECH, IMT Atlantique, Universit\'{e} de Nantes, CNRS-IN2P3, Nantes, France
\item \Idef{org115}Suranaree University of Technology, Nakhon Ratchasima, Thailand
\item \Idef{org116}Technical University of Ko\v{s}ice, Ko\v{s}ice, Slovakia
\item \Idef{org117}Technical University of Split FESB, Split, Croatia
\item \Idef{org118}The Henryk Niewodniczanski Institute of Nuclear Physics, Polish Academy of Sciences, Cracow, Poland
\item \Idef{org119}The University of Texas at Austin, Physics Department, Austin, Texas, United States
\item \Idef{org120}Universidad Aut\'{o}noma de Sinaloa, Culiac\'{a}n, Mexico
\item \Idef{org121}Universidade de S\~{a}o Paulo (USP), S\~{a}o Paulo, Brazil
\item \Idef{org122}Universidade Estadual de Campinas (UNICAMP), Campinas, Brazil
\item \Idef{org123}Universidade Federal do ABC, Santo Andre, Brazil
\item \Idef{org124}University of Houston, Houston, Texas, United States
\item \Idef{org125}University of Jyv\"{a}skyl\"{a}, Jyv\"{a}skyl\"{a}, Finland
\item \Idef{org126}University of Liverpool, Liverpool, United Kingdom
\item \Idef{org127}University of Tennessee, Knoxville, Tennessee, United States
\item \Idef{org128}University of the Witwatersrand, Johannesburg, South Africa
\item \Idef{org129}University of Tokyo, Tokyo, Japan
\item \Idef{org130}University of Tsukuba, Tsukuba, Japan
\item \Idef{org131}Universit\'{e} Clermont Auvergne, CNRS/IN2P3, LPC, Clermont-Ferrand, France
\item \Idef{org132}Universit\'{e} de Lyon, Universit\'{e} Lyon 1, CNRS/IN2P3, IPN-Lyon, Villeurbanne, Lyon, France
\item \Idef{org133}Universit\'{e} de Strasbourg, CNRS, IPHC UMR 7178, F-67000 Strasbourg, France, Strasbourg, France
\item \Idef{org134}Universit\`{a} degli Studi di Pavia, Pavia, Italy
\item \Idef{org135}Universit\`{a} di Brescia, Brescia, Italy
\item \Idef{org136}V.~Fock Institute for Physics, St. Petersburg State University, St. Petersburg, Russia
\item \Idef{org137}Variable Energy Cyclotron Centre, Kolkata, India
\item \Idef{org138}Warsaw University of Technology, Warsaw, Poland
\item \Idef{org139}Wayne State University, Detroit, Michigan, United States
\item \Idef{org140}Wigner Research Centre for Physics, Hungarian Academy of Sciences, Budapest, Hungary
\item \Idef{org141}Yale University, New Haven, Connecticut, United States
\item \Idef{org142}Yonsei University, Seoul, Republic of Korea
\item \Idef{org143}Zentrum f\"{u}r Technologietransfer und Telekommunikation (ZTT), Fachhochschule Worms, Worms, Germany
\end{Authlist}
\endgroup
  %%%%%%% done by webmaster team
%\input{}              %%%%%%%%%%%% put the references here
%
\end{document}